\documentclass[times]{cpeauth}
\usepackage{cite}
\usepackage{amsmath}
\usepackage{multirow}
\usepackage{enumerate}
\usepackage{graphicx}
\usepackage{csvsimple}
\usepackage{algorithm}
\usepackage{algcompatible}
\usepackage{multicol}
\usepackage{url}
\usepackage{float}

\usepackage{lscape}
\usepackage{longtable}

\usepackage{xcolor}
\usepackage{fancyhdr}

\DeclareMathOperator*{\maximize}{\textbf{maximize}}

\usepackage{afterpage}

\usepackage{subfig}

\usepackage{moreverb}

\usepackage[colorlinks,bookmarksopen,bookmarksnumbered,citecolor=red,urlcolor=red]{hyperref}

\newcommand\BibTeX{{\rmfamily B\kern-.05em \textsc{i\kern-.025em b}\kern-.08em
T\kern-.1667em\lower.7ex\hbox{E}\kern-.125emX}}

\def\volumeyear{2016}

\begin{document}

\runningheads{M H Ferdaus et al.}{Multi-objective, Decentralized Dynamic Virtual Machine Consolidation}

\title{Multi-objective, Decentralized Dynamic Virtual Machine Consolidation using ACO Metaheuristic in Computing Clouds}

\author{
	Md Hasanul Ferdaus\affil{1,3,}\corrauth, 
	Manzur Murshed\affil{2},
	Rodrigo N. Calheiros\affil{3},
	Rajkumar Buyya\affil{3}
}

\address{
	\affilnum{1} Faculty of Information Technology, Monash University, Clayton, VIC, Australia.\break
	\affilnum{2} Faculty of Science and Technology, Federation University Australia, Northways Road, Churchill, VIC 3842, Australia.\break
	\affilnum{3} Cloud Computing and Distributed Systems (CLOUDS) Laboratory, Department of Computing and Information Systems, Building 168, The University of Melbourne, Parkville, VIC 3053, Australia.
}

\corraddr{Md Hasanul Ferdaus, Faculty of Information Technology, Monash University, Clayton, VIC, Australia. Email: md.ferdaus@monash.edu}


\begin{abstract}
Underutilization of computing resources and high power consumption are two primary challenges in the domain of Cloud resource management. This paper deals with these challenges through offline, migration impact-aware, multi-objective dynamic Virtual Machine (VM) consolidation in the context of large-scale virtualized data center environments. The problem is formulated as an NP-hard discrete combinatorial optimization problem with simultaneous objectives of minimizing resource wastage, power consumption, and the associated VM migration overhead. Since dynamic VM consolidation through VM live migrations have negative impacts on hosted applications performance and data center components, a VM live migration overhead estimation technique is proposed, which takes into account pragmatic migration parameters and overhead factors. In order to tackle scalability issues, a hierarchical, decentralized dynamic VM consolidation framework is presented that helps to localize migration-related network traffic and reduce network cost. Moreover, a multi-objective, dynamic VM consolidation algorithm is proposed by utilizing the Ant Colony Optimization (ACO) metaheuristic, with integration of the proposed VM migration overhead estimation technique. Comprehensive performance evaluation makes it evident that the proposed dynamic VM consolidation approach outpaces the state-of-the-art offline, migration-aware dynamic VM consolidation algorithm across all performance metrics by reducing the overall power consumption by up to 47\%, resource wastage by up to 64\%, and migration overhead by up to 83\%. \newline
\end{abstract}

\keywords{Cloud Computing; Virtual Machine; Dynamic VM Consolidation; Energy Efficient; Migration Overhead; Migration Cost; Optimization; Data Center}

\maketitle

\vspace{-6pt}

\section{Introduction}
\label{chap5-introduction}
Cloud Computing paradigm provides access to computing resources and application services as a pay-as-you-go business model. Technically, Clouds are large pools of easily accessible and readily usable virtualized resources that can be dynamically reconfigured to adjust to a variable load via elasticity, and load balancing, and thus, offer opportunities for optimal resource utilization. This pool of virtualized resources is typically provisioned by Cloud infrastructure providers with extremely high availability and almost perfect reliability (e.g., 99.997\% for Amazon EC2 \cite{Leopold2015}) by means of Service Level Agreements (SLAs). Cloud consumers can access these resources and services based on their requirements without any regard as to the location of the consumed resources and services. 
In order to cope with the rapid growth of customer demands for processing power, storage, and communication, Cloud providers, such as Amazon, Google, and Microsoft are deploying large-scale data centers across the globe. Recent report shows that Cloud giant Amazon operates at least 30 data centers in its global network, each comprising 50,000 to 80,000 servers with a power consumption of between 25 to 30 megawatts \cite{Miller2015}. As a consequence, a huge amount of electrical energy is required to run the servers and network devices, as well as to keep the cooling systems operating for these data centers. 
In spite of continuous progress in equipment efficiency, statistics of the worldwide data center electricity consumption show non-linear growth throughout the last decade and a similar trend is expected for the upcoming years \cite{Mills2013}: a steady rise of 110\% from 2010 to 2015 and a predicted rise of 82\% from 2015 to 2020.
Large data centers are not only expensive to maintain, but they also have enormous detrimental effects on the environment. Reports claim that the information technology ecosystem alone represents around 10\% of the world's electricity consumption \cite{Clark2013} and data centers, the main driving element of this ecosystem, are responsible for around 2\% of global Greenhouse Gas (GHG) emissions, a share comparable to the aviation industry \cite{Vaughan2015}.

This extremely high energy consumption is not just because of the amount of computing resources used and the power inefficiency of the hardware infrastructures, but also due to the inefficient use of these resources. A recent study presented by Reiss \textit{et al.} \cite{Reiss2012} shows that a 12,000-nodes Google cluster achieves aggregate CPU utilization only of 25-35\% and memory utilization of 40\%. A similar underutilization trend has been identified by the researchers from Stanford University showing that a thousand-nodes production cluster at Twitter runs consistently at CPU utilization below 20\% and memory usage at around 40-50\%, whereas the overall utilization estimates are even poorer (between 6\% and 12\%) for Cloud facilities that do not consider workload co-location \cite{Delimitrou2014}. Moreover, the narrow dynamic power range of physical servers further exacerbates the problem--- even completely idle servers consume about 70\% of their peak power usage \cite{Fan2007}. Such low resource utilization, technically termed \textit{Server Sprawl}, contributes to both capital expenses and operational costs due to non-proportional energy consumption. As a consequence, underutilization of data center resources is a major challenge for the ultimate success of Cloud Computing.

\subsection{Motivation}
Cloud infrastructures depend on one or more data centers, either centralized or distributed, and on the use of various cutting-edge resource virtualization technologies that enable the same physical resources (computing, storage, and network) to be shared among multiple application environments \cite{Zhang2010}. 
Virtualization technologies allow data centers to address resource and energy inefficiencies by (i) provisioning multiple Virtual Machines (VMs) in a single physical server, where each VM represents a run-time environment completely isolated from one another, and (ii) live migrations of VMs \cite{Clark2005} from current hosts to other servers; and by this process, providing opportunities to improve resource utilization. In particular, efficient VM placement and consolidation decisions during the VM life cycle offer potential for the optimization of data center resources and power consumption

While online or on-demand VM placement and allocation techniques, such as the one presented in our previous work \cite{Ferdaus2014a}, have potentials to optimize placement decisions at the time of VM initiation, active VMs exhibit variations in actual resource usage during the VM life cycle due to workload variations. Furthermore, due to the features of on-demand resource provisioning and a pay-per-use business model, VMs are created and terminated dynamically and, as a consequence, data center resources become fragmented, which leads to degradation of server resource utilization and overall hosting capacity of the data center. Moreover, due to the narrow dynamic power range of physical servers, underutilized servers cause non-proportional power consumption in data centers. Both the problems of run-time server resource wastage and power consumption can be addressed by improving server resource utilization through the application of dynamic VM consolidation mechanism.

\textit{Dynamic VM Consolidation} focuses on run-time environments where VMs are active and already hosted by servers in the data center. Consolidation of such VMs is achieved by the VM live migration operations \cite{Clark2005, Nelson2005}, where a running VM is relocated from its current host to another server while it is still running and providing service to its consumers \cite{Vogels2008}. After the consolidation of the VMs, servers that are released by this process are turned to lower power states, such as standby or turned off, in order to save energy. Moreover, such consolidation improves the overall resource utilization of the active servers and resource wastage is minimized. Besides these obvious advantages, dynamic VM consolidation has other benefits such as data center physical space minimization, maintenance automation, and reduced labor costs. Nonetheless, this adds additional complexity to the consolidation operation since the current placement of the VMs needs to be considered while VM migration decisions are made. This is because VM migration operations incur migration impact or cost on both the hosted applications and data center components, such as the host server and communication network \cite{Verma2011}. As a consequence, any efficient dynamic VM consolidation must consider both the gain achieved by consolidating the VMs into a reduced number of servers and the overhead of the necessary VM migrations needed to achieve the consolidation. Therefore, such dynamic VM consolidation techniques need to be \textit{multi-objective}, where they opt for maximizing the gain of energy saving and resource utilization, as well as reducing the cost or overhead of necessary VM migrations.

Contrary to the existing methods mentioned above, the approach presented in this paper considers various migration parameters relating to VMs and data center components in order to estimate extent of realistic migration overheads. Such migration parameters and overheads are adopted from the insights demonstrated by previous measurement studies \cite{Akoush2010, Liu2013, Verma2010} on VM live migration in order to ensure the estimation technique is as pragmatic as possible. Such a realistic measure has obvious benefits over simplistic measures, such as the number of migrations, in that it can reveal the VM migrations that are beneficial for containing the migration overhead to a limited extent while at the same time improving server resource utilization. This is important since, in a dynamic data center environment, there can be instances where two or more VM migrations may have lower migration overhead than a different individual VM migration, an occurrence which it cannot be determined with a simplistic measure that considers only the number of migrations. 

Moreover, the migration overhead estimation method is further integrated with the proposed multi-objective, dynamic VM consolidation algorithm that generates migration plans for the VMs running in the data center. The primary benefit of this methodology is that it adopts a practical approach to quantifying the cost or impact of each VM migration in the data center that can be readily integrated with any other dynamic VM consolidation technique. Furthermore, unlike many of the existing studies \cite{Murtazaev2011, Marzolla2011, Beloglazov2012, Nguyen2014} that suggest greedy heuristics, the proposed dynamic VM consolidation technique adapts the multi-agent-based \textit{Ant Colony Optimization} (ACO) metaheuristic \cite{Dorigo1997} that works, in multiple iterations, based on solution refinement method. Utilization of such a refinement-based metaheuristic helps the algorithmic procedure to avoid early stagnation at local optima.

Furthermore, in order to address the scalability issue of data center-wide dynamic VM consolidation, a hierarchical, decentralized dynamic VM consolidation framework is proposed where the servers of the data center are grouped into clusters and it is recommended that VM consolidation operations be performed individually within the clusters. A network cost-aware cluster formation approach is suggested in this paper in order to localize the VM migration related network traffic within the lowest level of the network topology. This clustering approach has the advantage that migration-related network traffic does not travel through upper-layer network switches, and thereby helps avoiding data center network clogging. Moreover, such traffic does not need to travel long distances and therefore reduces the related network cost. Having said that, the proposed framework is not restricted to any cluster formation approach and any other dynamic clustering technique can be readily integrated with the framework, as well as with the dynamic VM consolidation algorithm.

The proposed techniques and strategies for multi-objective dynamic VM consolidation are built upon some assumptions regarding the data center environment. The migration overhead estimation technique assumes that the active memory size and page dirty rate of each VM running in the data center are known \textit{a priori}. In a virtualized data center, this information can be made readily available by utilizing virtualization Application Programming Interfaces (APIs), such as Red Hat \textit{libvirt}\footnote{libvirt: The virtualization API, 2016. \url{http://libvirt.org/}}, or using virtualization tools, such Red Hat \textit{virt tools}\footnote{virt tools: Open Source Virtualization Management Tools, 2016. \url{http://virt-tools.org/}}. It is further assumed that the inter-server network distance and available network bandwidth for performing VM live migrations are also known prior to the migration overhead estimation. Such network information can be measured using network tools, for example, the \textit{iPerf}\footnote{iPerf, 2016. \url{https://iperf.fr/}} network testing tool can be used to measure the maximum achievable bandwidth on IP networks, and the \textit{MTR}\footnote{MTR, 2016. \url{http://www.bitwizard.nl/mtr/}} network monitoring tool can measure the end-to-end network distance in terms of number of hops or in terms of delay in packet forwarding. 

Moreover, it is assumed that the hypervisors (e.g., Xen \cite{Barham2003} and KVM \cite{Kivity2007}) running in the data center servers are homogeneous. This is important to ensure the compatibility of the VM live migration operations among servers. In such an environment, it is presumed that several hypervisor properties relating to the VM live migration operation are already known, such as the remaining dirty memory threshold and the maximum number of rounds for the pre-copy migration. Furthermore, the dynamic VM consolidation algorithm takes into account the current resource demands of the active VMs in the data center, such as the CPU, main memory, and network I/O. In addition, the consolidation algorithm also needs to know the usable resource capacities of the server running in the data center. Last but not least, it is assumed that the data center network topology is already known for successful application of the proposed hierarchical, decentralized VM consolidation framework. 

\subsection{Contributions and Organization}
The \textbf{key contributions} of this paper are as follows:
\begin{enumerate}
\item The \textit{Multi-objective, Dynamic VM Consolidation Problem} (MDVCP) is formally defined as a discrete combinatorial optimization problem with the objective of minimizing data center resource wastage, power consumption, and overall migration overhead due to VM consolidation. 
\item VM migration overhead estimation models are proposed with consideration of realistic migration parameters and overhead factors in the context of the pre-copy VM live migration technique. The estimation models are not restricted to any specific VM consolidation method and can be readily integrated to any online or offline consolidation strategies. 
\item A hierarchical, decentralized VM consolidation framework is proposed to improve the scalability of dynamic VM consolidation in the context of medium to large-scale data centers. 
\item A novel \textit{ACO-based, Migration overhead-aware Dynamic VM Consolidation} (AMDVMC) algorithm is put forward as a solution to the proposed MDVCP problem. The AMDVMC algorithm
is integrated with the recommended decentralized consolidation framework and utilizes the proposed migration overhead estimation models. 
\item Extensive simulation-based experimentation and performance analysis is conducted across multiple scaling factors and several performance metrics. The results suggest that the proposed dynamic VM consolidation approach significantly optimizes the VM allocations by outperforming the compared migration-aware VM consolidation techniques across all performance metrics.
\end{enumerate}

The rest of this paper is organized as follows. The next section describes the closely related works in the area of VM consolidation. Section \ref{chap5-sec-dyn-vm-con} introduces the multi-objective, dynamic VM consolidation problem and presents the necessary mathematical frameworks to model it as a combinatorial optimization problem. Section \ref{chap5-sec-proposed-solution} describes the proposed VM live migration estimation models, the hierarchical, decentralized dynamic VM consolidation framework, and the ACO-based multi-objective, dynamic VM consolidation algorithm. Section \ref{chap5-sec-performance-evaluation} presents the performance evaluation and analysis of the results where the proposed dynamic VM consolidation approach is compared with other state-of-the-art approaches. Finally, Section \ref{chap5-sec-conclusions} concludes the paper with a summary of the contributions and results, as well as future research directions.

\section{Related Work}
\label{chap5-sec-related-work}
With the wide-spread use of various virtualization technologies and deployment of large-scale data centers, and most importantly with the advent of Cloud Computing, VM consolidation has emerged as an effective technique for data center resource management. As a consequence, a large number of research works have been carried out in the area of VM consolidation with different objectives, solution approaches, and system assumptions \cite{Ferdaus2014}. In terms of the necessary VM migrations, some of the works are migration impact-aware, whereas others do not consider the impact of VM migrations and therefore, are migration impact-unaware. 

VM consolidation is traditionally modeled as variants of bin packing or vector packing problems \cite{CoffmanJr1984}. The objective of such problems is to pack a group of VMs into a minimal number of servers so that the resource usage is reduced and eventually, become energy-efficient. Various greedy heuristic-based solution approaches are suggested for solving the migration-unaware VM consolidation problem, such as First Fit Decreasing (FFD) \cite{Verma2008, Wood2009}, Best Fit (BF) \cite{Mishra2011}, Best Fit Decreasing (BFD) \cite{Beloglazov2012a}, and so on \cite{Li2009, Li2012, Li2013}. Further works on greedy algorithm-based energy-aware VM placement approaches can be found in \cite{Srikantaiah2008} and \cite{Lim2009}. 

VM consolidation problem is also modeled as Constraint Satisfaction Problem (CSP). Based on the use of constraint solver, Hermenier et al. \cite{Hermenier2009} proposed Entropy, a dynamic server consolidation manager for clusters that finds solutions for VM consolidation with the goal of active server minimization and tries to find any reconfiguration plan of the proposed VM placement solution with objective to minimize the necessary VM migrations. Van et al. \cite{Van2010, NguyenVan2009} proposed VM provisioning and placement techniques to achieve high VM packing efficiency in Cloud data centers. However, a limitation of these approaches is that, by the use of constraint programming, the proposed solutions effectively restrict the domain of the total number of servers and VMs in data center and therefore, limit the search space and thus, suffer from scalability issues. 

Linear programming formulations are also applied for representing VM consolidation problems. Such linear programming formulations for server consolidation problems are presented in [45] and [6]. The authors also described constraints for limiting the number of VMs to be assigned to a single server and the total number of VM migrations, ensuring that some VMs are placed in different servers and placement of VMs to specific set of servers that has some unique properties. In order to minimize the cost of solving the linear programming problem, the authors further developed an LP-relaxation-based heuristic. Based on linear and quadratic programming model, Chaisiri et al. \cite{Chaisiri2009} presented an algorithm for finding optimal solutions to VM placement with the objective of minimizing the number of active servers. 

Since the ACO metaheuristics \cite{Dorigo2006} are proven to be effective in solving combinatorial optimization problems, specially for large-scale problem instances, several other works \cite{Feller2011, Gao2013, Ferdaus2014a} have proposed specialized static VM consolidation algorithms by utilizing the ACO metaheuristics, however these works focus on VM migrations where migration cost is not considered. Furthermore, evolutionary algorithms, such as genetic algorithms are also being applied in the problem domain of VM consolidation and a couple of such works can be found in \cite{Xu2010} and \cite{Mi2010}.

Several VM consolidation works addressed the problems of power inefficiency and cost of VM migrations  through multi-objective approaches. 
In \cite{Murtazaev2011}, the authors proposed Sercon, an iterative heuristic for dynamic VM consolidation that starts with a sorted list of PMs and VMs based on their current loads of CPU and memory. Thereafter, Sercon tries to migrate all the VMs from the least loaded PM to the most loaded one in the list. If unsuccessful, then it tries to migrate to the next least loaded PM. This process enforces the constraint that either all the VMs from one PM must be migrated to the other PM, or no migration happens. By this process, Sercon tries to minimize the number of VM migrations with the assumption that the least loaded PM will have least number of VMs. Marzolla \textit{et al.} \cite{Marzolla2011} proposed V-MAN, a decentralized, dynamic VM consolidation scheme where PMs coordinate with each other using a gossip-based message passing mechanism and VMs are migrated from lightly loaded PMs to heavily loaded ones. Each PM maintains a view of its neighbors and current VM allocation messages are dispatched among the members of a neighborhood. Upon receiving of allocation information, each PM tries to either push or pull some VMs from other neighbor PMs. However, V-MAN considers all VM having only a single resource of equal amount which is not a proper reflection of the VMs running in production data centers where each VM consumes several physical resources, such as CPU, memory, and network bandwidth. Furthermore, it considers only two PMs while making VM migration and consolidation decisions. Feller \textit{et al.} \cite{Feller2012} proposed a multi-objective, offline dynamic VM consolidation algorithm based on the Max-Min Ant System \cite{Stutzle2000} where the algorithm compute migration plans that reduces the number of active PMs, while at the same time keeping the number of VM migrations minimal. Similar work on dynamic VM consolidation can also be found in \cite{Farahnakian2015}. However, these approaches consider that every VM migration has equal migration overhead, however it is explained in the preceding section that this is an over-simplified assumption which is not pragmatic, specially in the context of dynamic data center environments, such as Clouds.

Most of the existing works on multi-objective dynamic VM consolidation mentioned above try to rearrange active VMs into the minimum number of servers in order to save energy, by turning the idle servers to lower power states while reducing the number of VM migrations needed to perform the VM consolidation operation \cite{Murtazaev2011, Marzolla2011, Feller2012, Beloglazov2012, Nguyen2014, Farahnakian2015}. A profound limitation that exists in these approaches is that every VM migration is considered equivalent in terms of migration cost or overhead. Experimental studies \cite{Verma2011, Voorsluys2009} on VM live migration shows that a migrating VM experiences performance degradation during its total migration time. For example, performance analysis on migrating a VM that hosts a web server reveals that the server throughput can drop by 20\% during the migration period \cite{Clark2005}. Moreover, a VM live migration operation results in a VM down period (formally, VM downtime) during which the services hosted by the VM remain unavailable to the consumers \cite{Nelson2005}. Furthermore, VM migration causes extra energy consumption for the servers and network devices, as well as generation of additional network traffic for transferring VM memory pages.

In an environment where different categories of applications are hosted by the running VMs, such as Cloud data centers, VMs exhibit high variations in their resource usage patterns and performance footprints. As a consequence, different VM migrations have different performance overhead both on the hosted services and on the data center components. A recent study \cite{Wu2015} has attempted to consider the VM migration cost while making consolidation decisions by incorporating a couple of VM properties into its consolidation scoring method. However, it overlooked data center properties, such as migration link bandwidth and network costs, as well as the dynamic nature of VM migration. Moreover, the evaluation is based solely on score values, rather than realistic properties. Therefore, dynamic VM consolidation approaches need to consider realistic measures of individual VM migration costs or overheads for making practical and efficient consolidation decisions.

\section{Multi-objective, Dynamic VM Consolidation Problem}
\label{chap5-sec-dyn-vm-con}
\begin{figure}[!t]
\centering
\includegraphics[scale=.5, trim=0.5cm 1.75cm 0cm 1.5cm]{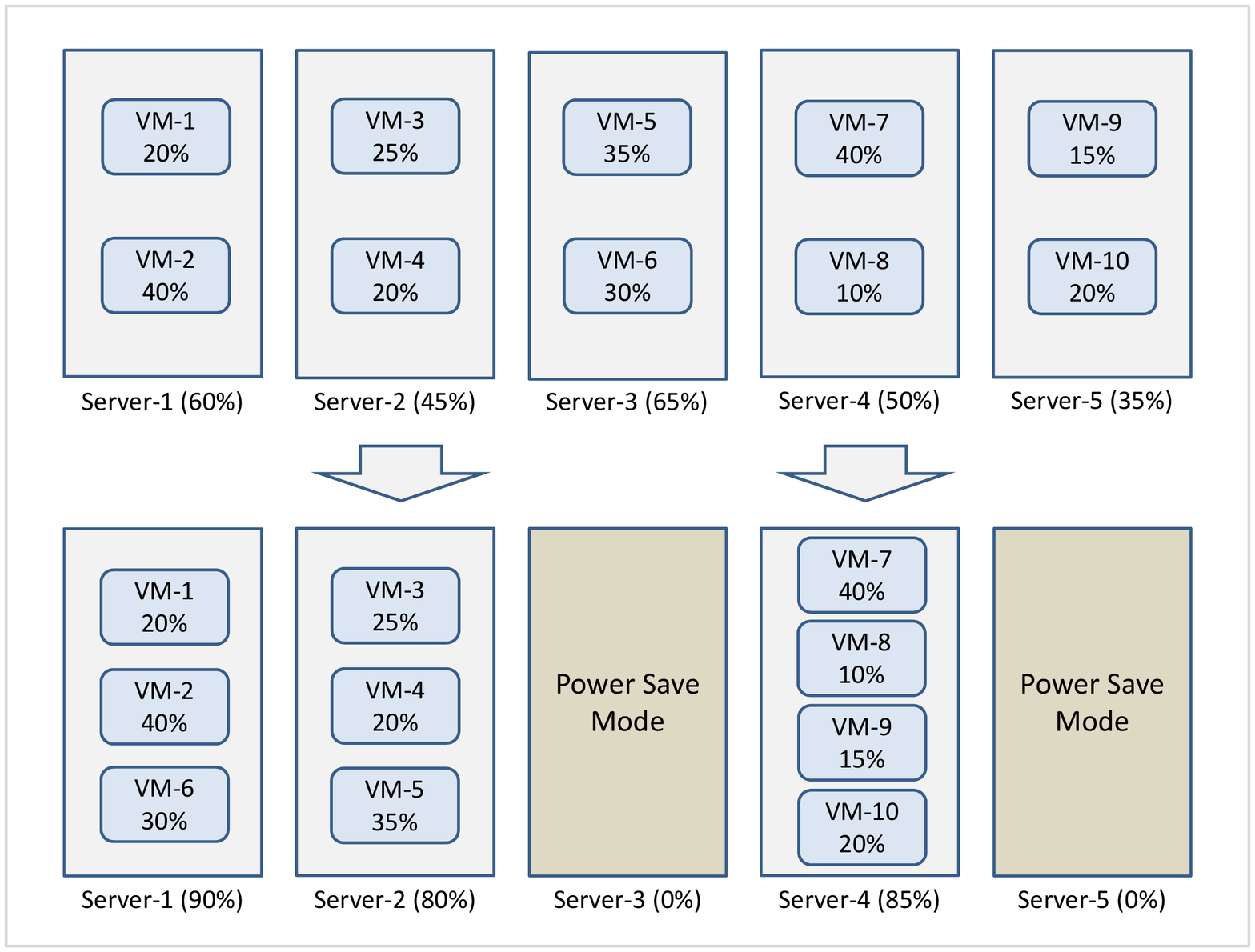}
\caption{Improving resource utilization and energy consumption through dynamic VM consolidation.}
\label{chap5-fig-dyn-vm-con-1}
\end{figure}

By the use of dynamic VM consolidation, active VMs are live migrated from one server to another to consolidate them into a minimal number of servers in order to improve overall resource utilization and reduce resource wastage. Servers released by this process can be turned to lower power states (such as suspended or turned off) with the goal of minimizing the overall power consumption. For example, in Figure \ref{chap5-fig-dyn-vm-con-1}, 10 VMs are running in 5 servers, each having an overall resource utilization of not more than 65\%. The VMs can be reassigned to be consolidated into 3 servers resulting in higher utilization and, by this process, 2 servers can be released and turned to power save mode in order to improve energy efficiency. 

However, dynamic VM consolidation at run-time is not merely a vector packing problem, such as the consolidated VM placement problem as presented in our previous work \cite{Ferdaus2014a}, since it needs to consider the current VM-to-server placements and the impact of necessary VM migrations on the performance of the hosted applications and the data center network \cite{Clark2005}. Current dynamic VM consolidation approaches \cite{Feller2012, Murtazaev2011, Marzolla2011} try to pack VMs into a minimal number of servers while reducing the number of migrations. However, the overhead of a live VM migration varies, based on some specific characteristics of the migrating VM and the corresponding network link used for migrating VM memory pages from the source server to the destination server. For example, a VM with 2 GB memory will be migrated faster than a VM with 4 GB memory, given that the other conditions are exactly same and thus, it needs less migration time and also requires fewer memory pages to be transferred from the source server to the destination server. For these reasons, merely considering the number of migrations needed for consolidation is an oversimplified metric to measure the impact of the necessary VM migrations. Therefore, in order to estimate the overhead or impact of the overall VM migrations required for achieving a particular VM consolidation state, it is important to estimate a realistic measure of the overhead of a single VM migration.

Given the above insights into the dynamic VM consolidation technique, the remaining part of this section formally defines the MDVCP problem as a discrete combinatorial optimization problem with necessary notations and models.

\subsection{Modeling Multi-objective, Dynamic VM Consolidation as a Combinatorial Optimization Problem}
\label{chap5-sec-modeling-amdvmc-com-opt}
Let $P\!M\!S$ denote the set of active servers or \textit{Physical Machines} (PMs) in a data center and $V\!M\!S$ denote the set of active VMs running on those PMs. $RC\!S$ represents the set of $d$ types of resources available in each PM. Table \ref{chap5-tab-notations-1} provides the various notations used in the problem definition and proposed solution. 
\begin{table}[!t]
\caption{Notations and their meanings}
\label{chap5-tab-notations-1}
\centering
\begin{tabular}{|l|l|}
\hline
\multicolumn{1}{|c|}{\text{$Notation$}} & \multicolumn{1}{|c|}{\text{$Meaning$}} \\
\hline
$v$			& 	Individual Virtual Machine	\\
$V\!M\!S$	& 	Set of active VMs in a data center	\\
$v^{cpu}$	& 	CPU demand of a VM $v$\\
$v^{mem}$	&	Memory demand of a VM $v$\\
$v^{dr}$	& 	Page Dirty Rate of a VM	\\
$v^{hp}$	& 	Host PM of a VM	\\
$N_{v}$		& 	Total number of VMs in a data center	\\
$V$			& 	Set of active VMs in a PM cluster	\\
$N_{vc}$	& 	Number of VMs in a PM cluster	\\
\hline

$p$			&	Individual Physical Machine	\\
$P\!M\!S$	& 	Set of active PMs in a data center	\\
$H\!V_{p}$	& 	Set of VMs hosted by PM $p$	\\
$N_{p}$		& 	Total number of PMs in a data center	\\
$P$			& 	Set of PMs in a PM cluster	\\
$N_{pc}$	& 	Number of PMs in a cluster	\\
\hline

$r$			& Single computing resource in PM (e.g., CPU, memory, network I/O)	\\
$RC\!S$		& Set of computing resources available in PMs	\\
$d$			& Number of resource types available in PM \\
\hline

$D\!S(p_{1}, p_{2})$	& Network distance between PMs $p_{1}$ and $p_{2}$	\\
$B\!A(p_{1}, p_{2})$	& Available bandwidth between PMs $p_{1}$ and $p_{2}$	\\
\hline

$OG(v,p)$		& Overall gain of assigning VM $v$ to PM $p$	\\
$U\!G_{p}(v)$ 	& Utilization gain of PM $p$ after VM $v$ is assigned in it \\
$M\!O(v,p)$ 	& Migration overhead incurred due to transferring VM $v$ to PM $p$	\\
$f$ 		& MDVCP Objective Function	\\
\hline

$M\!D$		& Amount of VM memory (data) transferred during a migration \\
$MT$ 		& Total time needed for carrying out a VM migration operation \\
$DT$ 		& Total duration during which VM is turned down during a migration \\
$N\!C$ 		& Network cost that will be incurred for a migration operation \\
\hline

$M\!EC$	& Energy consumption due to VM migration \\
$M\!SV$	& SLA violation due to VM migration \\
$M\!M$		& Migration map given by a VM consolidation decision \\
\hline
\end{tabular}
\end{table}

Each PM $p$ $(p \in P\!M\!S)$ has a d-dimensional \emph{Resource Capacity Vector} (RCV) $C_{p} = \{ C_{p}^{r} \}$, where $C_{p}^{r}$ denotes the total capacity of resource $r$ $(r \in RC\!S)$ of $p$. Similarly, each VM $v$ $(v \in V\!M\!S)$ is represented by its d-dimensional \emph{Resource Demand Vector} (RDV) $D_{v} = \{ D_{v}^{r} \}$, where $D_{v}^{r}$ denotes the demand of resource $r$ $(r \in RC\!S)$ of $v$. Moreover, memory page dirty rate and current host PM for a VM $v$ are denoted by $v^{dr}$ and $v^{hp}$, respectively.

The set of VMs hosted by a PM $p$ is denoted by $H\!V_{p}$. The \emph{Resource Utilization Vector} (RUV) of $p$ is denoted by $U_{p}=\{ U_{p}^{r} \}$, where $U_{p}^{r}$ denotes the utilization of resource $r$ $(r \in RC\!S)$ and is computed as the sum of the RDVs of its hosted VMs:
\begin{equation}
U_{p}^{r} = \sum_{v \in H\!V_{p}} D_{v}^{r}.
\label{chap5-eq-constr-pm-util-1}
\end{equation}

In this modeling, the data center is not restricted to any fixed network topology. Network distance and available network bandwidth used for VM live migration operations between any two PMs $p_{1}$ and $p_{2}$ are represented by $D\!S(p_{1},p_{2})$ and $B\!A(p_{1},p_{2})$. This network distance can be any practical measure, such as the number of hops or switches or network link latency in the communication path between $p_{1}$ and $p_{2}$. Thus, the network distance $D\!S$ and available bandwidth $B\!A$ models are generic and different model formulations focusing on any particular network topology or architecture can be readily applied in the optimization framework and proposed solution. Although singular distance between two PMs is considered here, link redundancy and multiple communication paths in data centers can be incorporated in the proposed model and the consolidation algorithm by appropriate definition of the distance function ($D\!S$) and the available bandwidth function ($B\!A$).

Given the above models and concepts, the objective of the MDVCP problem is to search for a VM migration decision for all the VMs in the data center that maximizes the number of released PMs (that can be turned to lower power states) at a minimal overall migration overhead, while respecting the PM resource capacity constraints. Therefore, the \textit{Objective Function} (OF) $f$ of the MDVCP problem can be expressed as follows:
\begin{equation}
\label{chap5-eq-f-1}
\maximize f(M\!M) = \frac{n\!ReleasedP\!M^{\phi}}{M\!O(M\!M)}
\end{equation}
where $M\!M$ is the \textit{Migration Map} for all the VMs in the data center which is defined as follows:
\begin{equation}
\label{chap5-eq-mm-1}
 M\!M_{v,p}=
   \begin{cases}
   1, & \text{if } V\!M \text{ } v \text{ is to be migrated to } P\!M \text{ } p; \\
   0, & \text{otherwise}.
  \end{cases}
\end{equation}
$M\!O(M\!M)$ represents the overall migration overhead of all the VM migrations denoted by migration map $M\!M$ which are necessary for achieving the consolidation and is expressed by Eq. \ref{chap5-eq-mo-2}. Details on measuring an estimation of the migration overhead ($M\!O(M\!M)$) is presented in the next section. And, $\phi$ is a parameter that signifies the relative importance between the number of released PMs ($n\!ReleasedP\!M$) and migration overhead ($M\!O$) for computing the OF $f$.

The above-mentioned OF is subject to the following PM resource capacity constraints:
\begin{equation}
\label{chap5-eq-pm-cap-constr-1}
\sum_{v \in V\!M\!S} D_{v}^{r}M\!M_{v,p} \leq C_{p}^{r}, \forall p \in P\!M\!S, \forall r \in RC\!S.
\end{equation}
The above constraint ensures that the resource demands of all the VMs that are migrated to any PM do not exceed PM's resource capacity for any of the individual resource types. And, the following constraint guarantees that a VM is migrated to exactly one PM:
\begin{equation}
\label{chap5-eq-mig-constr-1}
\sum_{p \in P\!M\!S}M\!M_{v,p} = 1, \forall v \in V\!M\!S.
\end{equation}

For a fixed number of PMs in a data center, maximization of the number of released PM ($n\!ReleasedP\!M$) otherwise means minimization of the number of active PMs ($n\!ActiveP\!M$) used for hosting the $N_{v}$ VMs. Moreover, minimization of the number of active PMs otherwise indicates reduction of the power consumption and resource wastage of the active PMs in a data center, as well as maximization of packing efficiency ($P\!E$). Thus, the above OF $f$ models the addressed MDVCP problem as a multi-objective problem. Moreover, it is worth noting that $f$ represents an expression of multiple objectives with potentially conflicting goals--- it is highly likely that maximization of the number of released PMs would require higher number of VM migrations, resulting in larger migration overhead. Therefore, any solution, to be efficient in solving the MDVCP problem, would require it to maximize the number of released PMs with minimal migration overhead.

Within the above definition, the MDVCP is represented as a discrete combinatorial optimization problem since the objective is to find a migration map (i.e., the optimal solution) from the finite set of all possible migration maps (i.e., solution space) that gives maximum value for the OF $f$. Furthermore, it is worth noting that the search space of the problem increases exponentially with $N_{v}$ and $N_{p}$. Effectively, the MDVCP problem falls in the category of $\mathcal{NP}\!\!-\!hard$ problem for which no exact solution can be obtained in realistic time. 


\section{Proposed Solution}			
\label{chap5-sec-proposed-solution}
A dynamic VM consolidation mechanism requires running VMs to be migrated and consolidated into fewer PMs so that empty PMs can be turned to lower power states in order to save energy. However, VM live migration impacts hosted applications, requires energy to transfer VM memory pages, and increases network traffic. Furthermore, these migration overheads vary from VM to VM, depending on several migration related parameters, such as VM memory size and the available bandwidth of the network link used for the migration. Therefore, dynamic VM consolidation schemes need to know a measure of the overhead for each VM migration in order to reduce the overall migration impact. 

In light of the above discussion, this section first presents a VM live migration overhead estimation model considering the relevant migration parameters. Secondly, in order to improve scalability of the dynamic VM consolidation algorithm and reduce network traffic incurred due to required VM migrations, a PM clustering scheme is presented that groups PMs in the data center based on the inter-PM network distances and dynamic consolidation being performed within each cluster locally. Finally, the migration overhead-aware dynamic VM consolidation algorithm, called AMDVMC, is proposed by utilizing the various models presented in this paper and in our previous paper \cite{Ferdaus2014a}.

\subsection{VM Live Migration Overhead Estimation}
\label{chap5-vm-mig-impact-measurement}
\emph{VM Live Migration} \cite{Clark2005} is a powerful feature of virtualization platforms that allows an active VM running live application services to be moved around within and across data centers. 
Nonetheless, VM live migration has a negative impact on the performance of applications running in a VM during the migration duration, on the underlying communication network due to the traffic resulting from transferring VM memory pages, as well as energy consumption due to carrying out the migration operation \cite{Liu2013}. These migration overheads can vary significantly for different application workloads due to the variety of VM configurations and workload patterns. For example, previous measurement studies on VM live migration demonstrated that VM downtime can vary significantly among workloads due to the differences in memory usage patterns, ranging from 60 milliseconds for a Quake 3 game server \cite{Clark2005} to 3 seconds in the case of high-performance computing benchmarks \cite{Nagarajan2007}. Another experimental study showed that applications hosted by migrating VMs suffer from performance degradation during the whole migration duration \cite{Voorsluys2009}. As a consequence, it is important to identify the relevant parameters that affect the migration process and the migration overhead factors that result from the process. To this end, the remaining part of this section presents a brief overview of the VM live migration process and details on the proposed migration overhead estimation models.


\subsubsection{Single VM Migration Overhead Estimation}
\label{chap5-single-vm-mig-overhead}
\begin{figure}[!t]
\centering
\includegraphics[scale=.5, trim=.5cm 9cm 0cm 1.5cm]{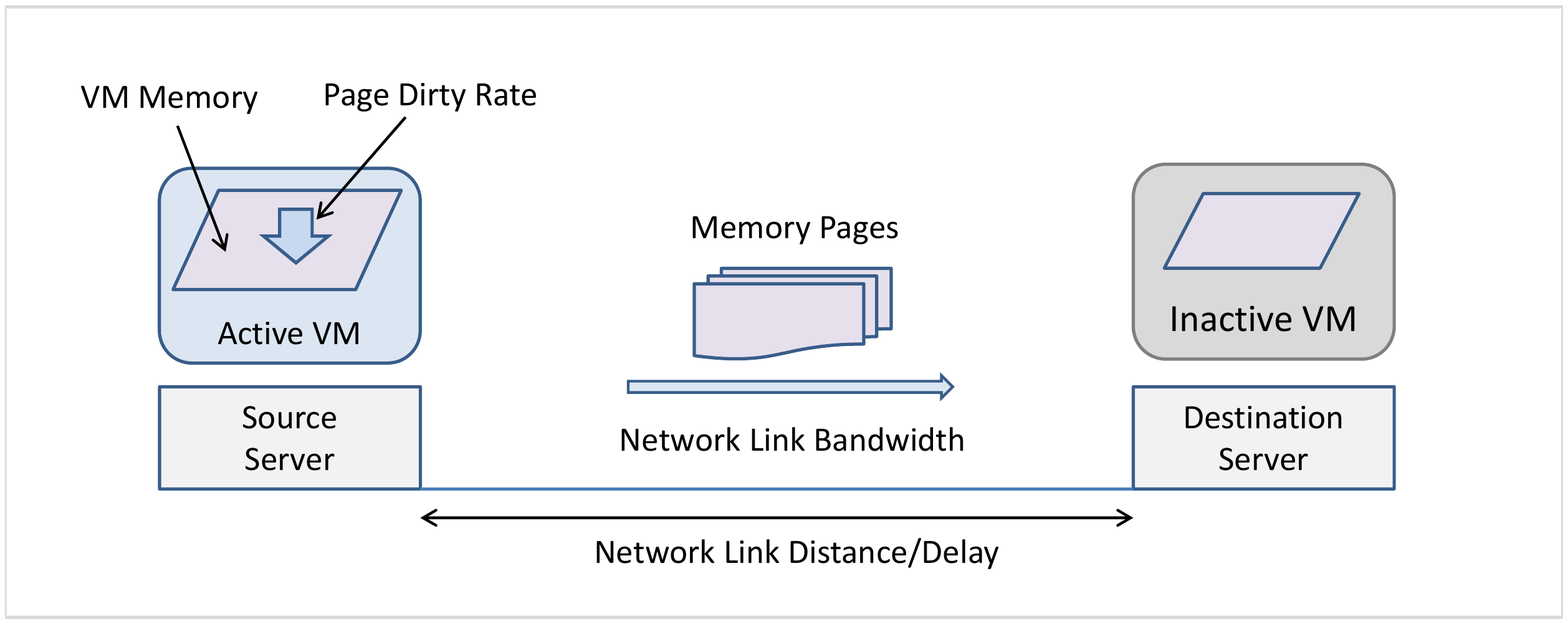}
\caption{Pre-copy VM live migration techniques and factors that affect migration overhead.}
\label{chap5-fig-precopy-mig-1}
\end{figure}

Among all the various VM live migration techniques, \emph{Pre-copy VM Migration} has been the most popular and widely used as the default VM migration subsystem in modern hypervisors, such as XenMotion \cite{Clark2005} for Xen Server and VMotion \cite{Nelson2005} for VMware ESXi Server. In this technique, the migrating VM continues to run in the source PM while the VM memory pages are iteratively transferred from the source to the destination (Figure \ref{chap5-fig-precopy-mig-1}). After a predefined number of iterations are completed, or a pre-specified amount of dirty memory remains to be transferred, or any other terminating condition is met, the VM is stopped at the source, the remaining memory pages are moved to the destination and, finally, the VM is restarted in the destination PM. The obvious benefits of this technique are the relatively short stop-and-copy phase and, therefore, shorter VM downtime compared to other live migration techniques, and higher reliability as it retains an up-to-date VM state in the source machine during the migration process. However, pre-copy migration can require longer total migration time since memory pages can be transmitted multiple times in several rounds depending on the page dirty rate and, for the same reason, it can generate much higher network traffic compared to other approaches. Therefore, given the migration parameters (e.g., VM memory size and page dirty rate) that affect VM live migration, it is necessary to estimate measures of migration overhead factors properly (e.g., total migration time and VM downtime) in order to decide which VMs to migrate for dynamic consolidation so as to reduce the overhead incurred due to the migration operation. 

\paragraph{VM Migration Parameters} There are several migration parameters that affect migration performance and, hence, the accuracy of migration overhead estimation \cite{Akoush2010}:

\begin{enumerate}
\item \textit{VM Memory Size ($v^{mem}$):} In the first iteration, the pre-copy migration scheme transfers the whole VM memory from the source PM to the destination PM and, thus, the duration of the first iteration is directly proportional to the memory size. As a result, the memory size impacts the total migration duration and, on average, this duration varies linearly with the memory size. Moreover, a larger memory indicates that more network traffic will be generated for performing the migration operation.

\item \textit{VM Page Dirty Rate ($v^{dr}$):} After the first iteration, only the memory pages that are modified (i.e., dirty) during an iteration are copied from the source to the destination PM in the next iteration. Thus, a higher page dirty rate causes more data transfer per iteration and results in a longer total migration duration. Furthermore, a higher page dirty rate indicates that more memory pages need to be transferred in the last transfer round and, as a consequence, this increases the VM downtime. 

\item \textit{Migration Link Bandwidth ($B\!A$):} For each of the pre-copy migration data transfer rounds, a higher link bandwidth will enable faster data transmission and shorten the round duration. As a result, both the total migration time and VM downtime will be reduced. Thus, the migration link bandwidth is inversely proportional to the total migration time and VM downtime.

\item \textit{Migration Link Distance ($D\!S$):} Since VM migration causes a non-negligible amount of data transfer through the communication network, it incurs a traffic overhead on the network links of the data center. In this context, the migration link distance can refer to the physical distance that migration-related data needs to be transferred or the latency/delay in data communication from the source PM to the destination PM. Therefore, the migration link distance has a direct effect on the overall network overhead for the migration.
\end{enumerate}

Apart from the above-mentioned general parameters, the VM migration overhead can vary based on two migration configuration parameters of the specific hypervisor:
\begin{enumerate}
\item Threshold value for the remaining amount of dirty memory ($DV_{th}$), and
\item Maximum number of rounds for the pre-copy migration algorithm ($max\_round$).
\end{enumerate}
When the pre-copy migration process reaches either of the above two points, the VM is stopped, the remaining dirty memory is transferred, and the VM is restarted in the destination PM (termed \textit{stop-and-copy} phase). However, for a data center with homogeneous hypervisors, these two parameters can be considered predefined and fixed and, therefore, these are considered as constants in the proposed overhead estimation model.

\paragraph{Migration Overhead Factors} Given the above migration parameters, the proposed model estimates the following four migration overhead factors that contribute to overall migration overhead:

\begin{enumerate}
\item \textit{Migration Data Transferred ($M\!D$):} As the pre-copy migration process involves multiple rounds for sending dirtied memory in the previous rounds from the source to the destination, the total amount of data transferred due to the migration can be equal to or greater than the VM memory size. This factor has a direct impact on the energy consumption and network overhead incurred due to the migration.

\item \textit{Migration Time ($MT$):} This implies the total duration for the migration from the initiation of the migration process to the point when the migrated VM has started running in the destination PM. This is an important element of the overall migration impact since the migrating VM suffers from performance degradation during the migration duration \cite{Beloglazov2012}.

\item \textit{VM Downtime ($DT$):} This represents the time duration for which the VM would be halted and the hosted service would be unavailable to the consumers. VM downtime is composed of the time duration needed for the stop-and-copy phase to transfer the remaining dirty memory in the last iteration and the time spent in resuming the VM at the destination PM.

\item \textit{Network Cost ($N\!C$):} The network cost or overhead of a VM migration is modeled as the product of the total data (i.e., memory pages) transferred ($M\!D$) from the source PM to the destination PM during the migration process and the network distance ($DS$) between PMs. Since data center networks are usually designed in a hierarchical fashion (e.g., tree topology \cite{Al-Fares2008}), VM migrations that involve transmission of migration data through higher layer network switches (e.g., core/aggregation switch) incur more network cost compared to migrations that involve data transmission among PMs under the same access switches. 
\end{enumerate}

The model for estimating single VM migration overhead follows the internal operational steps of the pre-copy migration technique and extends the VM live migration performance modeling presented by Liu \textit{et al.} \cite{Liu2013}. For the purpose of completeness, the algorithmic steps of the process (VMMigOverhead) is presented in Algorithm \ref{chap5-alg-vmmigoverhead-1}. As input, it takes the above-mentioned migration parameters and the destination PM, and computes estimates for the above-mentioned migration overhead factors. The algorithm starts by initializing $M\!D$ and $MT$ to zero which store the estimates of the total data to be transmitted and time duration for the whole migration process. 

After setting $p_{s}$ to the current host PM of the VM, the VMMigOverhead algorithm checks whether the source and destination PMs are the same [lines 1--2]. If yes, then it sets $DT$ and $N\!C$ to zero and terminates, since there is no memory data transfer in this case [lines 3--6].

If the source and destination PMs differ, it sets $DV_{0}$ to the VM memory size which indicates that the whole memory will be transmitted during the first round as per the pre-copy migration technique [line 7]. 

In each migration round [lines 8--19], the model estimates the time duration of the round ($T_{i}$) by dividing the memory data to be transferred ($DV_{i}$) in this round (which is estimated in the previous round) by the available bandwidth ($B\!A$) of the migration network link [line 9]. It also estimates the size of the Writable Working Set (WWS) for the next round and deducts it from the memory size that is expected to be dirtied in this round in order to estimate the memory data that will be transmitted in the next round [lines 10--12]. The WWS is deducted since it represents the memory pages that are modified very frequently and are skipped during the pre-copy rounds. And, $\mu_{1}$, $\mu_{2}$, and $\mu_{3}$ are model parameters that can be learned through benchmarking and learning techniques (e.g., linear regression). Table \ref{chap5-tab-mig-parameters-1} shows the values of these parameters used for the purpose of performance evaluation. Details on this derivation and parameter values can be found in the original paper \cite{Liu2013}. 

If the estimate of the memory data to be transferred in the next round goes below the predefined threshold ($DV_{th}$) or is greater than the memory data that is estimated to be transferred in this round [line 13], then it indicates that the termination condition is met and the next round would be the stop-and-copy phase of the migration process. For that round, it estimates the size of the memory data to be transferred, the duration of the stop-and-copy phase, and the VM downtime ($DT$) [lines 14--16]. Finally, the algorithm estimates the total memory data to be transferred ($M\!D$) and the migration duration ($MT$) by accumulating the memory data size and the time duration for each of the rounds, respectively [lines 20--23], as well as the network cost as a product of the total memory data and the network distance between the VM's current host PM ($p_{s}$) and the destination PM ($p$) [line 24].
\begin{algorithm}[!t]
\textbf{Input:} $v^{mem}$, $v^{dr}$, $B\!A$, $D\!S$, and $p$. \\
\textbf{Output:} $M\!D$, $MT$, $DT$, and $N\!C$. \\
\textbf{Initialization:} $M\!D \leftarrow 0$; $MT \leftarrow 0$. \\

\begin{algorithmic}[1]
\STATE	$p_s \leftarrow v^{hp}$
\IF{$p_{s} = p$} \COMMENT{Check whether the source PM and destination PM are the same}
	\STATE	$DT \leftarrow 0$
	\STATE	$N\!C \leftarrow 0$
	\STATE 	\textbf{return}
\ENDIF	

\STATE	$DV_{0} \leftarrow v^{mem}$ \COMMENT{In the first iteration, the whole VM memory is transferred}

\FOR{$i=0$ to $max\_round$}
	\STATE	$T_{i} \leftarrow DV_{i}/B\!A(p_{s}, p)$ \COMMENT{Estimate the time duration for this pre-copy round}
	\STATE	$\kappa \leftarrow \mu_{1} \times T_{i} + \mu_{2} \times v^{dr} + \mu_{3}$
	\STATE	$W_{i+1} \leftarrow \kappa \times T_{i} \times v^{dr}$ \COMMENT{Estimate the size of WWS for the next round}
	\STATE	$DV_{i+1} \leftarrow T_{i} \times v^{dr} - W_{i+1}$ \COMMENT{Estimate the migration data size for the next round}

	\IF{$DV_{i+1} \leq DV_{th} \vee DV_{i+1} > DV_{i}$} \COMMENT{Check if termination condition is met}
		\STATE	$DV_{i+1} \leftarrow T_{i} \times v^{dr}$
		\STATE	$T_{i+1} \leftarrow DV_{i+1}/B\!A(p_{s}, p)$
		\STATE 	$DT \leftarrow T_{i+1} + T_{res}$ \COMMENT{Estimate the duration of VM downtime}
		\STATE 	\textbf{break}
	\ENDIF	
\ENDFOR

\FOR{$i=0$ to $max\_round$}
	\STATE $M\!D \leftarrow M\!D + DV_{i}$ \COMMENT{Estimate the total memory data transfer}
	\STATE $MT \leftarrow MT + T_{i}$    \COMMENT{Estimate the total migration time}
\ENDFOR

\STATE $N\!C \leftarrow M\!D \times D\!S(p_{s}, p)$ \COMMENT{Estimate network cost for the migration}

\end{algorithmic}
\caption{VMMigOverhead Algorithm}
\label{chap5-alg-vmmigoverhead-1}
\end{algorithm}

Finally, the unified \emph{Migration Overhead} $M\!O$ for migrating a VM $v$ from its current host ($v^{hp}$) to the destination PM $p$ is modeled as a weighted summation of the estimates of the above-mentioned migration overhead factors computed by algorithm VMMigOverhead:
\begin{equation}
\label{chap5-eq-mo-1}
M\!O(v,p) = \alpha_{1} \times M\!D(v,p) + \alpha_{2} \times MT(v,p) + \alpha_{3} \times DT(v,p) + \alpha_{4} \times N\!C(v,p)
\end{equation}
where $\alpha_{1}$, $\alpha_{2}$, $\alpha_{3}$, and $\alpha_{4}$ are input parameters that indicate the relative importance of the contributing migration overheads and $\alpha_{1}, \alpha_{2}, \alpha_{3}, \alpha_{4} \in [0,1]$ such that $\sum_{i=0}^4 \alpha_{i} = 1$. In order to keep the migration overhead within a fixed range of $[0,1]$ so that it is compatible to be integrated into dynamic VM consolidation mechanisms, all the contributory factors $M\!D$, $MT$, $DT$, and $N\!C$ are normalized against their maximum possible values before feeding them to compute the migration overhead $M\!O$.

\subsubsection{Modeling Energy Consumption due to Migration}
\label{chap5-model-energy-con-mig}
In a large data center with hundreds or thousands of running VMs, dynamic VM consolidation decisions can involve a large number of VM migrations. As a result, energy consumption due to the migration decisions should be taken into account and migration decisions that require a lower amount of energy consumption should be given preference over those that require higher energy. Since VM live migration is an I/O intensive task, the energy is primarily consumed due to the memory data transfer from the source to the destination. This data transfer involves the  source PM, the destination PM, and the network switches. This work utilizes the migration energy consumption model presented by Liu \textit{et al.} \cite{Liu2013}. Here, the energy consumption by the switches is not taken into account due to the inherent complexity of the switching fabric. Moreover, since the amount of data transmitted by the source PM and the amount of data received by the destination PM are equal, it can be assumed that the energy consumption by the two ends are the same. Therefore, the migration energy consumption due to a single VM migration is shown to be linearly correlated with the amount of memory data transmitted from the source to the destination:
\begin{equation}
\label{chap5-eq-mec-1}
M\!EC(v,p) = \gamma_{1} \times M\!D(v,p) + \gamma_{2}
\end{equation}
where $\gamma_{1}$ and $\gamma_{2}$ are model parameters. The specific values of these parameters used for the evaluation are shown in Table \ref{chap5-tab-mig-parameters-1} and are taken as reported by Liu \textit{et al.} \cite{Liu2013}. When migration data ($M\!D$) is measured in Megabytes, the migration energy consumption ($M\!EC$) is estimated in Joules.

\subsubsection{Modeling SLA Violation due to Migration}
\label{chap5-model-sla-vio-mig}
Since the applications hosted by a migrating VM experience performance degradation during the course of the migration operation, it is important to estimate the corresponding SLA violation and take this into consideration for any VM consolidation operation \cite{Beloglazov2012}. Obviously, VM consolidation decisions that result in fewer SLA violations compared to others are preferable in terms of migration overhead. An experimental study \cite{Voorsluys2009} on the impact of VM migration on an application demonstrates that performance degradation depends on the application behavior, particularly on the number of memory pages modified by the application (i.e., page dirty rate). Furthermore, the study suggests that, for the class of web-applications, the average performance degradation can be estimated at around 10\% of the CPU utilization during the migration operation. Therefore, the SLA violation due to a VM migration is modeled as follows:
\begin{equation}
\label{chap5-eq-msv-1}
M\!SV(v,p) = \sigma \times v^{cpu} \times MT(v,p)
\end{equation}
where $\sigma$ is an input parameter that indicates the percentage of performance degradation due to VM migration.

\subsubsection{Overall VM Migration Overhead due to Dynamic VM Consolidation}
For medium to large data centers, an offline, dynamic VM consolidation operation can require multiple VM migrations in order to achieve the desired consolidation. With the single VM migration overhead estimation models presented above, the estimates of the aggregated VM migration overhead factors are defined below:
\begin{enumerate}
\item Each VM migration data ($M\!D$) implies the amount of VM memory data that is needed to be transferred from the source PM to the destination PM and this data amount is directly proportional to the amount of energy needed for performing the migration operation. Therefore, the estimate of the aggregated migration data that will be transferred due to the VM migrations represented by a particular migration map $M\!M$ is given by:
\begin{equation}
\label{chap5-eq-md-1}
M\!D(M\!M) = \sum_{\langle v,p \rangle \in M\!M} M\!D(v,p).
\end{equation}
\item Since the applications, which are hosted by a VM, experience performance degradation during the period of VM migration ($MT$) and therefore, the corresponding SLA violation is proportional to the migration time, the aggregated migration time for all the VM migrations represented by migration map $MM$ is modeled as follows:
\begin{equation}
\label{chap5-eq-mt-1}
MT(M\!M) = \sum_{\langle v,p \rangle \in M\!M} MT(v,p).
\end{equation}
\item For any VM performing live migration, the services provided by the hosted applications remain unavailable to the consumers during the total period of the VM's downtime. In order to reflect on the overall service outage of the migrating VMs, the aggregated VM downtime is measured as the accumulated downtime of all the migrating VMs given by migration map $M\!M$:
\begin{equation}
\label{chap5-eq-dt-1}
DT(M\!M) = \sum_{\langle v,p \rangle \in M\!M} DT(v,p).
\end{equation}
\item The network cost ($N\!C$) that is incurred due to each VM migration implies the amount of additional network traffic and the corresponding energy consumption by the network switches due to the migration operation. Therefore, the network costs are considered additive and the aggregated network cost for a particular migration map $M\!M$ is given by:
\begin{equation}
\label{chap5-eq-nc-1}
N\!C(M\!M) = \sum_{\langle v,p \rangle \in M\!M} N\!C(v,p).
\end{equation}
\end{enumerate}
Given the above estimation models for single VM migration overhead factors, the overall migration overhead for a particular dynamic VM consolidation plan (or $M\!M$) for a group or cluster of PMs is modeled as the accumulated overhead of all the necessary VM migrations within that group or cluster:
\begin{equation}
\label{chap5-eq-mo-2}
M\!O(M\!M) = \sum_{\langle v,p \rangle \in M\!M} M\!O(v,p).
\end{equation}
Similarly, the estimate of the aggregated migration energy consumption for migration map $M\!M$ is computed as the summation of the migration energy consumption of the individual VMs:
\begin{equation}
\label{chap5-eq-mec-2}
M\!EC(M\!M) = \sum_{\langle v,p \rangle \in M\!M} M\!EC(v,p).
\end{equation}
And, the estimate of aggregated SLA violation for all the VM migrations given by a migration map $M\!M$ is defined as follows:
\begin{equation}
\label{chap5-eq-msv-2}
M\!SV(M\!M) = \sum_{\langle v,p \rangle \in M\!M}M\!SV(v,p).
\end{equation}

\subsection{Hierarchical, Decentralized, Dynamic VM Consolidation Framework}
\label{chap5-sec-hierar-dec-dvmc-frame}
\begin{figure}[!t]
\centering
\includegraphics[scale=.6, trim=3.5cm 9.5cm 0cm 1.5cm]{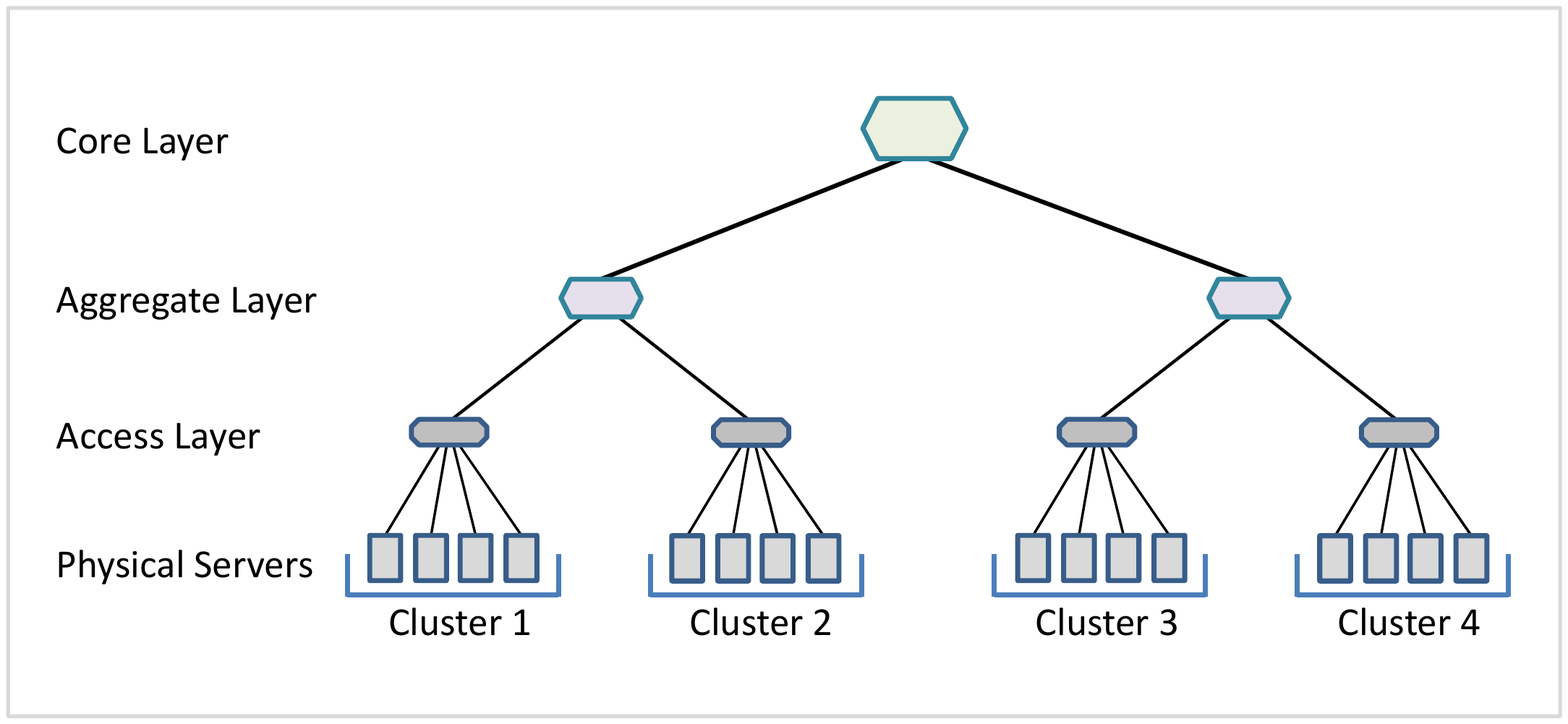}
\caption{Clustering data center servers based on network proximity.}
\label{chap5-fig-server-clusters-1}
\end{figure}

This subsection presents a hierarchical, decentralized, dynamic VM consolidation framework. In order to achieve scalability, PMs in a data center are grouped into smaller clusters and the dynamic VM consolidation operation is performed separately within each cluster. With the goal of reducing the network overhead incurred due to the necessary VM migrations resulting from any dynamic VM consolidation operation, PM clusters are formed based on the network cost of data communications among the PMs. The network cost can be derived through practical measures, such as the ones presented in \cite{Shrivastava2011, Meng2010, Korupolu2009}. In this proposed framework, the number of switches in the data communication path among the PMs is considered as a measure of the network distance and, based on this definition, PMs under the same access switch are grouped as an individual cluster (Figure \ref{chap5-fig-server-clusters-1}). However, such hierarchical, decentralized framework and the VM consolidation algorithm are not restricted to this particular cluster formation approach and any other static or dynamic cluster formation techniques can be readily integrated with this framework. Cluster formation based on network proximity ensures that VMs are migrated to short distant target servers and this then limits the overall migration impact on hosted applications and data center network. This is reflected in the network cost incurred due to the migration decision.

\begin{figure}[!t]
\centering
\includegraphics[scale=.47, trim=0cm 0cm 0cm .5cm]{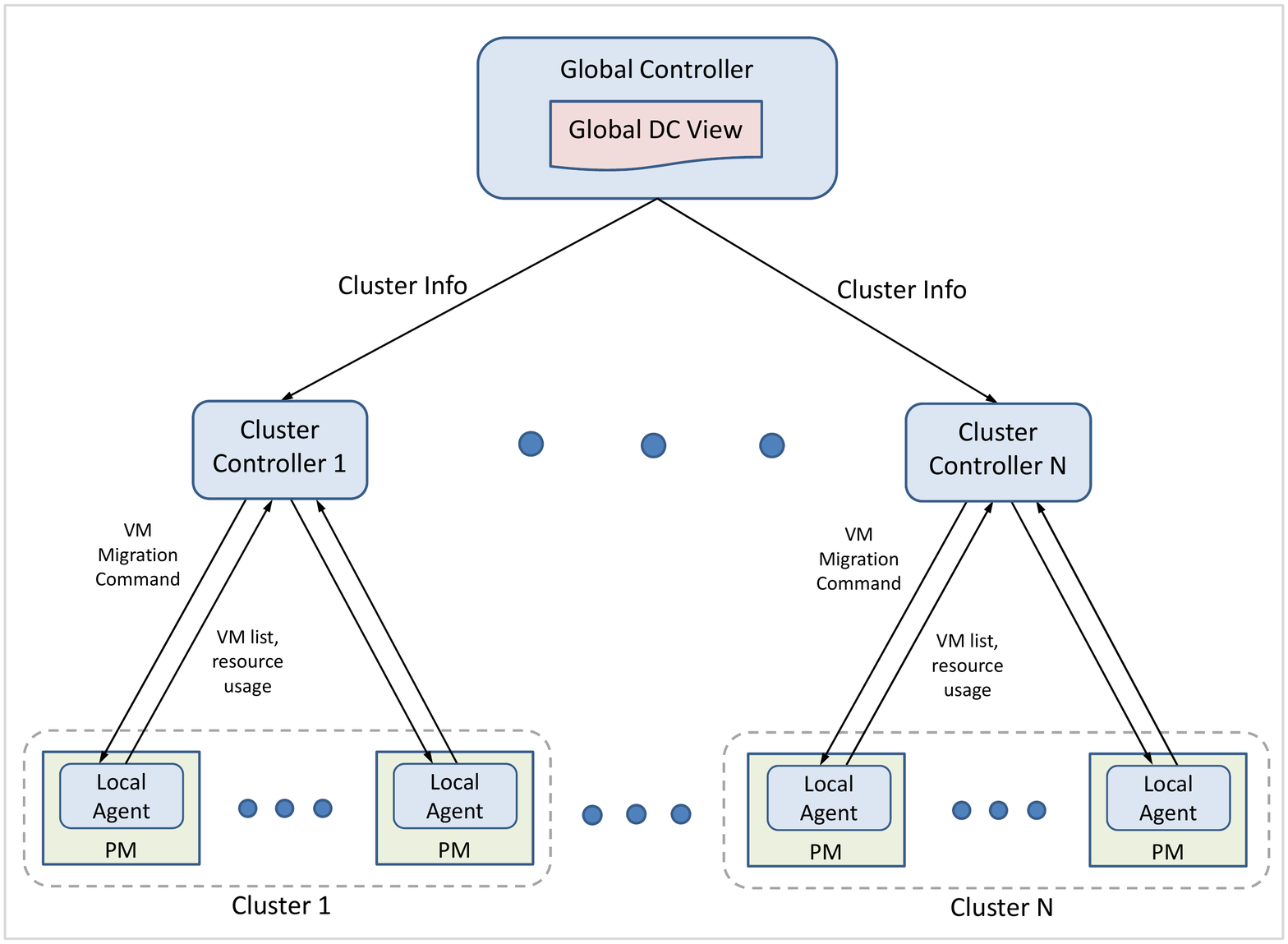}
\caption{A Hierarchical, Decentralized Dynamic VM Consolidation Framework.}
\label{chap5-fig-dec-dvmc-framework-1}
\end{figure}
Figure \ref{chap5-fig-dec-dvmc-framework-1} presents an overview of the hierarchical structure of the framework. Each PM in a cluster runs a Local Agent that collects VM related information, such as a list of hosted VMs ($H\!V_{p}$) and their resource usage ($D_{v}$). The Global Controller is the topmost entity that has a global view of the data center, including the list of PM ($P\!M\!S$) and network information ($D\!S$ and $B\!A$), and is responsible for cluster formation decisions. The Global Controller periodically sends cluster related information to each of the Cluster Controllers, such as a set/list of PMs in a cluster ($P$). Within each cluster, the Cluster Controller periodically receives information of the hosted VMs from each of the Local Agents and forms a cluster-wide list of VMs ($V$) hosted by the PMs. Within each cluster, the Cluster Controller periodically receives information on the hosted VMs from each of the Local Agents and forms a cluster-wide list of VMs ($V$) hosted by the PMs. When a triggering event occurs for the offline, dynamic VM consolidation operation (e.g., periodic or resource utilization threshold-based), each Cluster Controller runs the dynamic VM consolidation algorithm for its cluster and issues the necessary VM migration commands to the respective hypervisors. Global Controller and Cluster Controllers selection decision can be made either using static configuration or dynamic cluster leader selection algorithms \cite{Lo2005a, Dong2009}. However, this aspect is beyond the scope of this paper.

\subsection{Migration Overhead-aware, Multi-objective Dynamic VM Consolidation Algorithm}
\label{chap5-multi-obj-mi-aware-dvmc-alg}
\begin{figure}[!t]
\centering
\includegraphics[scale=.7, trim=4.3cm 7.5cm 0cm .8cm]{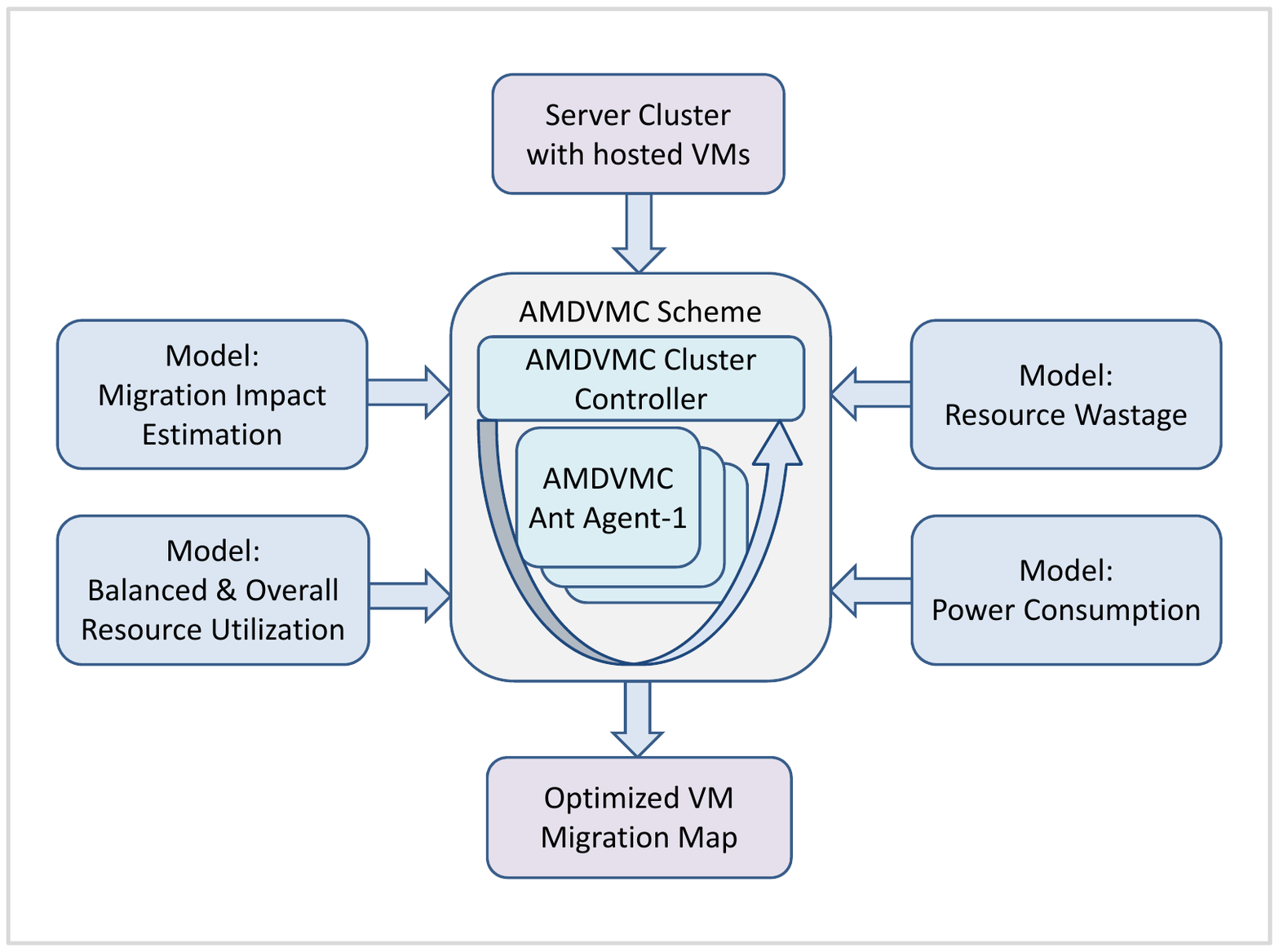}
\caption{AMDVMC algorithm with associated models.}
\label{chap5-fig-amdvmc-alg-1}
\end{figure}

This subsection presents the migration overhead-aware, multi-objective dynamic VM consolidation algorithm (AMDVMC) based on the \textit{Ant Colony System} (ACS) metaheuristic \cite{Dorigo1997} that iteratively refines migration plans in order to maximize the OF $f$ (Eq. \ref{chap5-eq-f-1}). The ACO-based metaheuristic is chosen as a solution for the MDVCP problem because of its proven effectiveness in the field of combinatorial optimization and polynomial time complexity \cite{Dorigo2003}. The main components of the proposed AMDVMC algorithm are shown in Figure \ref{chap5-fig-amdvmc-alg-1}. As input, it takes the PM cluster along with the hosted VMs, and the AMDVMC consolidation scheme makes use of the single and overall VM migration overhead models presented in Section \ref{chap5-vm-mig-impact-measurement}, as well as the resource wastage and power consumption models presented in our previous work \cite{Ferdaus2014a}. Within the scheme, the AMDVMC Controller creates multiple ant agents and delivers every ant an instance of the input PM cluster. The ants run in parallel, compute solutions (i.e., VM migration maps $M\!M=\{\langle v,p \rangle\}$), and pass the maps to the Controller. Each migration map consists of a list of VM-to-server migration commands ($\langle v,p \rangle$) for all the VMs in the cluster. For the migration commands where the source and the destination PMs are the same, all the migration factors and overhead for  these VMs would be zero and would not contribute to the overall migration overhead. The AMDVMC Controller then detects the best migration map based on the OF $f$ (Eq. \ref{chap5-eq-f-1}), updates the shared pheromone information, and executes the ants once again for the next cycle. Finally, when the predefined stop condition is met, the controller outputs the so-far-found best migration map.

\subsubsection{Adaptation of ACO Metaheuristic for AMDVMC}
\label{chap5-amdvmc-aco-adap}
Following a similar adaption performed for the AVVMC algorithm proposed in our previous work \cite{Ferdaus2014a} for solving the consolidated VM cluster placement problem (CVPP), each VM-to-PM migration within a cluster is considered as an individual solution component for adapting the ACS metaheuristic \cite{Dorigo1997} in order to solve the multi-objective dynamic VM consolidation problem. However, from the perspective of the solution-building process of ACS metaheuristic, there are two fundamental differences between the consolidated VM cluster placement problem and the MDVCP problem defined in this paper:
\begin{enumerate}
\item The initial data center states of the CVPP and MDVCP are not the same: in the case of CVPP, VMs are initially unassigned and the PMs are considered empty whereas in the case of the MDVCP, VMs are already assigned to their host PMs and therefore the PMs are not empty.
\item In the case of CVPP, each VM-to-PM assignment provides some benefit in terms of resource utilization but does not have any associated overhead or negative impact, whereas for the MDVCP, migrating a VM to a PM other than its current host provides some benefit in terms of resource utilization, but at the same time, incurs migration overhead.
\end{enumerate}

Without any loss of generality in the ACS's solution-building process, the first difference is addressed by considering that the VMs are initially pulled out of the host PMs and kept in a virtual VM pool and, in this way, the PMs can be considered virtually empty. As for the second difference, both the OF $f$ (Eq. \ref{chap5-eq-f-1}) and the heuristic information (Eq. \ref{chap5-eq-heuristic-info-1}) are updated in order to reflect the differences. 

Pheromone values are associated to each VM-to-PM migration that denotes the desirability of migrating a VM to a target PM (Eq. \ref{chap5-eq-init-pheromone-1} \& Eq. \ref{chap5-eq-global-pher-update-1}) and are implemented using an $N_{v} \times N_{p}$ pheromone matrix $\tau$. During the solution-building process, heuristic values are computed dynamically for each VM-to-PM migration, which represents the preference of migrating a VM to a target PM in terms of both PM resource utilization and VM migration overhead (Eq. \ref{chap5-eq-heuristic-info-1}). In an ACO cycle, each ant agent generates a solution (migration map) comprising of a list of VM-to-PM migrations. At the end of each cycle, the best solution is identified based on the OF $f$ (Eq. \ref{chap5-eq-f-1}) value and the pheromone levels of the solution components are updated so as to navigate the search space more effectively and shun early stagnation to a sub-optimal.

\subsubsection{The AMDVMC Algorithm}
\label{chap5-sec-amdvmc-alg}
The pseudocode of the proposed AMDVMC algorithm is shown in Algorithm \ref{chap5-alg-amdvmc-pseudocode-1}. It starts with a list of PMs ($P$) in a cluster along with the set of hosted VMs ($V$) and the relevant parameters as input, and generates a migration map ($M\!M$) as output. At the beginning of each cycle, each ant starts with an empty migration map, a set of empty PMs having total resource capacities similar to the PMs in $P$ (generated by subroutine $EmptyP\!M\!Set$), and a set of VMs having total resource demands similar to the VMs in $V$ (generated by subroutine $CopyV\!M\!Set$), and shuffles the VMs in $vm\!List$ [lines 2--7]. Ants work with empty PMs and the VMs are considered to be not-yet-placed in order to facilitate consolidation of VMs with the goal of maximizing resource utilization and eventually, minimizing energy consumption by increasing the number of released PMs. Moreover, when assigning VMs to PMs, ants take into consideration where the VMs are currently hosted and the corresponding migration overhead is taken into account for making the migration decisions. And, shuffling the VMs in $vm\!List$ adds randomization in the subsequent search process. 

Within lines 11--21, all the ants generate their migration maps (solutions) using a modified ACS rule (Eq. \ref{chap5-eq-pseudo-random-rule-1}). In each while loop iteration, an ant is chosen randomly [line 12]. If the ant has at least one VM in its $vm\!List$, it chooses a VM-to-PM migration from all the feasible VM migration options for the VM in $vm\!List$, adds the $\langle V\!M,\!P\!M \rangle$ pair in its migration map ($M\!M$), and removes the VM from its  $vm\!List$ [lines 13--16]. Otherwise, the ant has finished making migration decisions for all the VMs and it computes the OF ($f$) value for its solution according to (Eq. \ref{chap5-eq-f-1}) and the ant is removed from $antList$ [lines 17--20]. 

When all the ants have completed building their solutions, the while loop ends and the new \textit{Global-best Migration Map} ($G\!B\!M\!M$) is identified by comparing the existing $G\!B\!M\!M$ with the newly computed migration maps [lines 23--29]. Thereafter, the pheromone reinforcement amount is computed based on the quality of the $G\!B\!M\!M$ [line 31] accordingly to (Eq. \ref{chap5-eq-global-pher-update-2}) and the pheromone matrix is updated by simulating pheromone evaporation and deposition for each $\langle V\!M,\!P\!M \rangle$ pair accordingly to (Eq. \ref{chap5-eq-global-pher-update-1}) [lines 32--36]. The algorithm reinforces the pheromone value only on the $\langle V\!M,\!P\!M \rangle$ pairs that belong to the $G\!B\!M\!M$.

Finally, the algorithm checks whether there has not been any improvement in the quality of the solution for the last $nCycleTerm$ cycles or a total of $nResetMax$ cycle resets have occurred [line 37]. If it finds improvement, the search process repeats; otherwise, the algorithm terminates with the current $G\!B\!M\!M$ as output. The $nResetMax$ parameter is used to set an upper bound on the number of cycle resets so that AMDVMC does not run indefinitely. The remainder of this section formally defines the various parts of the AMDVMC algorithm.
\begin{algorithm}[!h]
\small
\textbf{Input:} Set of PMs $P$ and set of VMs $V$ in the cluster, set of ants $antSet$. Set of parameters $\lbrace nAnts, nCycleTerm, nResetMax, \omega ,\lambda, \beta, \delta, q_{0}, a, b \rbrace$. \\
\textbf{Output:} Global-best migration map $G\!B\!M\!M$. \\
\textbf{Initialization:} Set parameter values, set pheromone value for each $\langle V\!M\!,P\!M \rangle$ pair ($\tau_{v,p}$) to $\tau_{0}$ [Eq. \ref{chap5-eq-init-pheromone-1}], $G\!B\!M\!M \leftarrow \emptyset, nCycle \leftarrow 0, nCycleReset \leftarrow 0$. \\
\begin{algorithmic}[1]
\REPEAT

	\FOR{\textbf{each} $ant \in antSet$} \COMMENT{Initialize data structures for each ant}
		\STATE 	$ant.mm \leftarrow \emptyset$
		\STATE 	$ant.pm\!List \leftarrow Empty\!P\!M\!Set(P)$
		\STATE 	$ant.vm\!List \leftarrow CopyV\!M\!Set(V)$
		\STATE Shuffle $ant.vm\!List$ \COMMENT{Shuffle VMs to randomize search}
	\ENDFOR
\STATE

	\STATE $nCycle \leftarrow nCycle + 1$
	\STATE	$antList \leftarrow antSet$
	\WHILE{$antList \neq \emptyset$}
		\STATE Pick an $ant$ randomly from $antList$
		
		\IF{$ant.vm\!List \neq \emptyset$}
			\STATE Choose a $\langle v,p \rangle$ from set $\{\langle v,p \rangle | v \in ant.vm\!List, p \in ant.pm\!List\}$ according to (Eq. \ref{chap5-eq-pseudo-random-rule-1})		
			\STATE $ant.mm \leftarrow ant.mm \cup \langle v,p \rangle$ \COMMENT{Add the selected $\langle v,p \rangle$ to the ant's migration map}
			\STATE	$ant.vm\!List.remove(v)$
		\ELSE \COMMENT{When all VMs are placed, then ant completes a solution and stops for this cycle}
			\STATE	Compute the objective function (OF) value for $ant.mm.f$ according to (Eq. \ref{chap5-eq-f-1})
			\STATE	$antList.remove(ant)$
		\ENDIF
	\ENDWHILE
\STATE

	\FOR{\textbf{each} $ant \in antSet$} \COMMENT{Find global-best migration map for this cycle}
		\IF{$ant.mm.f > GBMM.f$}
			\STATE 	$GBMM \leftarrow ant.mm$
			\STATE 	$nCycle \leftarrow 0$
			\STATE  $nCycleReset \leftarrow nCycleReset + 1$
		\ENDIF
	\ENDFOR
\STATE

	\STATE	Compute $\Delta\tau$ based on (Eq. \ref{chap5-eq-global-pher-update-2}) \COMMENT{Compute pheromone reinforcement amount for this cycle}
	\FOR{\textbf{each} $p \in P$} \COMMENT{Simulate pheromone evaporation and deposition for this cycle}
		\FOR{\textbf{each} $v \in V$}
			\STATE $\tau_{v,p} \leftarrow (1-\delta) \times \tau_{v,p} + \delta \times \Delta\tau_{v,p}$
		\ENDFOR
	\ENDFOR
		
\UNTIL{$nCycle = nCycleTerm$ \textbf{ or } $nCycleReset = nResetMax$} \COMMENT{AMDVMC ends either if it sees no progress for consecutive \textit{nCycleTerm} cycles, or a total of \textit{nResetMax} cycle resets have taken place}
\end{algorithmic}
\caption{AMDVMC Algorithm}
\label{chap5-alg-amdvmc-pseudocode-1}
\end{algorithm}

\paragraph{Definition of Pheromone and Initial Pheromone Amount:} 
ACO algorithms \cite{Dorigo2006} start with a fixed amount of pheromone value for each of the solution components. For each solution component (here each $\langle v,p \rangle$ migration pair), its pheromone level provides a measure of desirability for choosing it during the solution-building process. In the context of AMDVMC, a fixed and uniform pheromone level for each of the solution components means that, at the beginning, each VM-to-PM migration has equal desirability. Following the approach used in the original ACS metaheuristic \cite{Dorigo1997}, the initial pheromone amount for AMDVMC is set to the quality of the migration map generated by the referenced L1 norm-based First Fit Decreasing (FFDL1) baseline algorithm:
\begin{equation}
\label{chap5-eq-init-pheromone-1}
\tau_{0} \leftarrow f_{F\!F\!D\!L\!1}.
\end{equation}

\paragraph{Definition of Heuristic Information:}
Heuristic value provides a measure of preference for selecting a solution component among all the feasible solution components during the solution-building process. For the AMDVMC algorithm, heuristic value $\eta_{v,p}$ indicates the apparent benefit of migrating a VM $v$ to a PM $p$ in terms of the improvement in the PM's resource utilization and the overhead incurred for migrating $v$ to $p$. However, an increase in a PM's resource utilization provides a positive incentive for improving the quality of the overall migration decision, whereas the migration overhead works as a negative impact since it reduces the quality of the migration decision according to the OF $f$ (Eq. \ref{chap5-eq-f-1}). Therefore, the heuristic value $\eta_{v,p}$ for selecting $\langle v,p \rangle$ migration is measured as follows:
\begin{equation}
\label{chap5-eq-heuristic-info-1}
\eta_{v,p} = \lambda \times U\!G_{p}(v) + (1-\lambda) \times (1 - M\!O(v,p))
\end{equation}
where $U\!G_{p}(v)$ is the \textit{utilization gain} of PM $p$ after placing VM $v$ in it and is computed as follows:
\begin{equation}
\label{chap5-eq-ug-2}
U\!G_{p}(v) = \omega \times (-\text{log}_{10} \|R\!I\!V_{p}(v)\|) + (1-\omega) \times U\!tilization_{p}(v)
\end{equation}
where $\|R\!I\!V_{p}(v)\|$ is the magnitude of the \textit{Resource Imbalance Vector} (RIV) of the PM $p$ after assigning the VM $v$ to it (defined by Eq. 7 in \cite{Ferdaus2014a}), $U\!tilization_{p}(v)$ is the overall resource utilization of the PM $p$ if the VM $v$ is assigned to it (defined by Eq. 14 in \cite{Ferdaus2014a}), and $\omega \in [0,1]$ is a parameter that trades off the relative importance of balanced versus overall resource utilization; and $M\!O(v,p)$ is the migration overhead incurred due to transferring the VM $v$ to the PM $p$ as expressed in Eq. \ref{chap5-eq-mo-1}. Finally, $\lambda \in [0,1]$ is a parameter that sets the relative weight between the achieved utilization gain and migration overhead incurred as per the definition. In order to ensure metric compatibility for the heuristic formulation (Eq. \ref{chap5-eq-heuristic-info-1}), both the utilization gain $U\!G$ and migration overhead $M\!O$ are normalized against their maximum values.

\paragraph{Pseudo-random Proportional Rule:}
During the migration map generation process (Algorithm \ref{chap5-alg-amdvmc-pseudocode-1}, line 14), an ant $k$ uses the following probabilistic decision rule \cite{Dorigo1997} to select a VM $v$ to be migrated to PM $p$:
\begin{equation}
\label{chap5-eq-pseudo-random-rule-1}
 s=
  \begin{cases}
   \text{arg } \text{max}_{v \in FM_{k}(vm\!List,pm\!List)} \lbrace \tau_{v,p} \times [\eta_{v,p}]^\beta \rbrace, & \text{if } q \leq q_{0}; \\
   S, & \text{otherwise}	
  \end{cases}
\end{equation}
where $q \in [0,1]$ is a uniform random number, $q_{0} \in [0,1]$ is an input parameter, $\eta_{v,p}$ is the heuristic value for $\langle v,p \rangle $ migration (Eq. \ref{chap5-eq-heuristic-info-1}), $\tau_{v,p}$ is the current pheromone value of ${\langle}{v,p}{\rangle}$ pair (Eq. \ref{chap5-eq-global-pher-update-1}), $\beta$ is a non-negative parameter that trades off between the significance of the pheromone amount and the heuristic value in the decision rule, and $S$ is a random variable selected according to the probability distribution given below by (Eq. \ref{chap5-eq-random-rule-1}). $FM_{k}(vm\!List,pm\!List)$ defines the set of feasible migrations ($\langle v,p \rangle$) for ant $k$ based on the VMs in $vm\!List$ and PMs in $pm\!List$ (i.e., VM migrations that do not violate the resource capacity constraint of target PM $p$ given by Eq. \ref{chap5-eq-pm-cap-constr-1}):
\begin{equation}
\label{chap5-eq-fm-1}
FM_{k}(vm\!List,pm\!List) = \left\{ \langle v,p \rangle \middle | \forall l \in \textit{RCS}, \forall v \in vm\!List, \forall p \in pm\!List: U_{p}^{l}+D_{v}^{l} \leq C_{p}^{l} \right\}.
\end{equation}
The above-mentioned decision rule works as follows: when $q \leq q_{0}$, then the ${\langle}{v,p}{\rangle}$ pair that results in the largest $\tau_{v,p} \times [\eta_{v,p}]^\beta$ value is selected and added to the migration map (exploitation), otherwise a ${\langle}{v,p}{\rangle}$ pair is chosen with probability $P_{k}(v,p)$ using the following \textit{random-proportional rule} (exploration):
\begin{equation}
\label{chap5-eq-random-rule-1}
 P_{k}(v,p)=
  \begin{cases}
   \frac{\tau_{v,p} \times [\eta_{v,p}]^\beta}{ \sum_{\langle u,p \rangle  \in FM_{k}(vm\!List,pm\!List)} \tau_{u,p} \times [\eta_{u,p}]^\beta}, & \text{if } \langle u,p \rangle \in FM_{k}(vm\!List,pm\!List); \\
    0, & \text{otherwise.}	
  \end{cases}
\end{equation}
The above random-proportional rule uses the pheromone values ($\tau_{v,p}$) of each $\langle v,p \rangle$ pair multiplied by the corresponding heuristic value ($\eta_{v,p}$) so as to prefer $\langle v,p \rangle$ pairs that improve PM resource utilization (both balanced and overall) and incur lower migration overhead, as well as having larger pheromone values. 

\paragraph{Global Pheromone Update:}
With the aim of favoring the VM-to-PM migrations that constitute the GBMM so that the ants can be better guided in the following iterations, the pheromone level of each $\langle v,p \rangle$ pair is updated using the following rule:
\begin{equation}
\label{chap5-eq-global-pher-update-1}
\tau_{v,p} \leftarrow (1-\delta) \times \tau_{v,p} + \delta \times \Delta\tau_{v,p}
\end{equation}
where $\delta$ is the global pheromone decay parameter ($0<\delta<1$) and $\Delta\tau_{v,p}$ is the pheromone reinforcement applied to each ${\langle}{v,p}{\rangle}$ pair that make up the GBMM. The value of the reinforcement is measured based on the quality of the solution in terms of the OF ($f$) value:
\begin{equation}
\label{chap5-eq-global-pher-update-2}
\Delta\tau_{v,p} = 
	\begin{cases}	
		f(G\!B\!M\!M), & \text{if } {\langle}{v,p}{\rangle} \in \textit{GBMM};\\
		0, & \text{otherwise.}
	\end{cases}
\end{equation}

\section{Performance Evaluation}
\label{chap5-sec-performance-evaluation}
This section presents the performance evaluation of the proposed AMDVMC algorithm through simulation-based experimentation where the results are compared to both migration impact-unaware and migration impact-aware dynamic VM consolidation algorithms.

\subsection{Algorithms Compared}
\label{chap5-sec-alg-compared}
The following VM consolidation algorithms are implemented and compared:

\subsubsection{First Fit Decreasing based on L1-norm (FFFL1)}
The FFFL1 algorithm is used as the baseline algorithm for the performance evaluation. This algorithm does not take into account the current VM-to-PM placements and it is, therefore, a migration impact-unaware algorithm. Scalability is ensured by running FFDL1 separately for each PM cluster as presented in the previous subsection. For each cluster, VMs are considered to be pooled out of the PMs and sorted in decreasing order of their resource demands. The L1-norm mean estimator is utilized to represent the three different resources (CPU, memory, and network I/O) into a scalar form. Thereafter, FFDL1 places each VM from the sorted list in the first feasible PM in the cluster following the \textit{First Fit} (FF) approach. The VM placements are subject to the resource capacity constraints represented by Eq.\ref{chap5-eq-pm-cap-constr-1} \& Eq. \ref{chap5-eq-mig-constr-1}. When the dynamic consolidation is performed for all the PM clusters, data center-wide performance metrics are accumulated: migration overhead-related factors using Eq. \ref{chap5-eq-md-1}-\ref{chap5-eq-msv-2}, and resource wastage and power consumption using Eq. 8-10 of our previous work \cite{Ferdaus2014a}. 

\paragraph{Time Complexity:} For the above-mentioned implementation, the worst-case time complexity for the FFDL1 algorithm is given by:
\begin{equation}
\small
\label{chap5-eq-ffdl1-time-complx-1}
\begin{aligned}
T_{\!F\!F\!D\!L1} = \mathcal{O}(N_{vc}\text{lg}N_{vc}) + \mathcal{O}(N_{vc}N_{pc}).
\end{aligned}
\end{equation}
For the cases of average Cloud data centers, $N_{pc}$ is expected to be greater than $\text{lg}N_{vc}$. With this assumption, the above equation can be reduced to the following:
\begin{equation}
\small
\label{chap5-eq-ffdl1-time-complx-2}
\begin{aligned}
T_{\!F\!F\!D\!L1} = \mathcal{O}(N_{vc}N_{pc}).
\end{aligned}
\end{equation}

\paragraph{Memory Overhead:} Given that merge sort \cite{Cormen2001} is used in FFDL1 implementation, then the memory overhead for sorting the VMs in a cluster would be $\mathcal{O}(N_{vc})$. Apart from sorting, the placement decision part of FFDL1 works in-place without using any additional data structure. Therefore, the overall memory overhead of the FFDL1 algorithm is given by: 
\begin{equation}
\small
\label{chap5-eq-ffdl1-mem-overhead-1}
\begin{aligned}
M_{\!F\!F\!D\!L1} = \mathcal{O}(N_{vc}).
\end{aligned}
\end{equation}

\subsubsection{Max-Min ant system-based Dynamic VM Consolidation (MMDVMC)}
The MMDVMC algorithm \cite{Feller2012} is an offline, dynamic VM consolidation algorithm that is executed in a random neighborhood of PMs based on an unstructured Peer-to-Peer network \cite{Voulgaris2005}. It aims to increase the number of released PMs and the variance of the scalar valued PM used capacity vectors, and reduce the number of necessary VM migrations within each neighborhood of the data center. It utilizes the Max-Min Ant System (MMAS) \cite{Stutzle2000} to solve the dynamic consolidation problem where multiple ant agents iteratively refine migration plans. This algorithm runs for $nCycles$ cycles and in each cycle a total of $nAnts$ ant agents compute solutions. In each cycle, every ant selects the VM migrations that eventually maximize the defined objective function value. At the end of each iteration, the cycle-best migration plan is determined and compared against the existing global-best plan in order to identify the new global-best migration plan. Finally, pheromone values are updated for each VM-PM pair using a pheromone update rule that bounds that pheromone values with a pre-defined range of [$\tau_{max}, \tau_{min}$]. The algorithm runs for a pre-defined number of iterations and returns that final global-best migration plan. The relevant parameter values for the algorithm are taken as reported in the original paper \cite{Feller2012}. The MMDVMC algorithm takes into account the number of VM migrations as a measure of migration overhead or impact, which is analyzed as an oversimplified measure in Section \ref{chap5-vm-mig-impact-measurement}. Similar to FFDL1, data center-wide performance metrics (resource, power, and migration overhead) are accumulated over all the clusters using the pre-defined formulations.

\paragraph{Time Complexity:} From the algorithmic pseudocode presented in the original paper \cite{Feller2012}, the worst-case time complexity of MMDVMC algorithm can be given by:
\begin{equation}
\small
\label{chap5-eq-mmdvmc-time-complx-1}
\begin{aligned}
T_{\!M\!M\!D\!V\!M\!C} 	&= \mathcal{O}(nCycles.nAnts.N_{vc}.N_{pc}.N_{vc}) \\
						&= \mathcal{O}(nCycles.nAnts.N_{vc}^{2}.N_{pc}).
\end{aligned}
\end{equation}
Since the optimal values of the parameters relating to the ACO metaheuristic are measured using preliminary experiments and they are not considered as a scaling factor for the VM consolidation problem, both $nCycles$ and $nAnts$ terms can be considered constant. Therefore, the above complexity can be simplified as follows:
\begin{equation}
\small
\label{chap5-eq-mmdvmc-time-complx-2}
\begin{aligned}
T_{\!M\!M\!D\!V\!M\!C} 	&= \mathcal{O}(N_{vc}^{2}.N_{pc})
\end{aligned}
\end{equation}

\paragraph{Memory Overhead:} Since MMDVMC has used the MMAS metaheuristic, it has a memory overhead of $\mathcal{O}(N_{vc}N_{pc})$ for maintaining pheromone information. Moreover, it has another $\mathcal{O}(nAnts)$ memory overhead for managing $nAnts$ ant agents. Furthermore, in every iteration, each ant agent computes its own migration plan, using its local list of PMs for a cluster with their associated hosted VMs, and modifies the VM-to-PM assignments. As a consequence, each ant agent has another $\mathcal{O}(N_{vc}N_{pc})$ memory overhead due to the local information of a cluster. Therefore, the overall memory overhead of MMDVMC is given by:
\begin{equation}
\small
\label{chap5-eq-mmdvmc-mem-overhead-1}
\begin{aligned}
M_{\!M\!M\!D\!V\!M\!C} 	&= \mathcal{O}(N_{vc}.N_{pc}) + \mathcal{O}(nAnts.N_{vc}.N_{pc}) \\
						&= \mathcal{O}(nAnts.N_{vc}.N_{pc}).
\end{aligned}
\end{equation}
Considering the number of ants is fixed, the memory overhead can be simplified as follows:  
\begin{equation}
\small
\label{chap5-eq-mmdvmc-mem-overhead-2}
\begin{aligned}
M_{\!M\!M\!D\!V\!M\!C} 	= \mathcal{O}(N_{vc}.N_{pc}).
\end{aligned}
\end{equation}

\subsubsection{ACO-based Migration impact-aware Dynamic VM Consolidation (AMDVMC)}
The proposed AMDVMC algorithm is implemented based on the description presented in Section \ref{chap5-sec-amdvmc-alg} and follows the execution flow presented in Algorithm \ref{chap5-alg-amdvmc-pseudocode-1}. The PMs in the data center are grouped into PM clusters, as presented in Section \ref{chap5-sec-hierar-dec-dvmc-frame}, and dynamic VM consolidation is performed by executing the AMDVMC algorithm in each cluster separately. Finally, data center-wide performance metrics (resource, power, and migration overhead) are accumulated over all the clusters using pre-defined formulations.
\paragraph{Time Complexity:} The time complexity for initializing the ant-related data structures (lines 2--7) is $T_{2\text{--}7} = \mathcal{O}(nAnts.N_{vc})$. The complexity of the migration map generation process for each of the ants is $\mathcal{O}(N_{vc}.N_{pc})$. Therefore, the time complexity of the migration maps generation for $nAnt$  (lines 11--21) would be $T_{11\text{--}21} = \mathcal{O}(nAnts.N_{vc}.N_{pc})$. Thereafter, the new GBMM identification part [lines 23--29] requires $T_{23\text{--}29} = \mathcal{O}(nAnts)$ time. Finally, the pheromone update part [lines 31--36] has $T_{31\text{--}36} = \mathcal{O}(N_{vc}.N_{pc})$ time complexity. Therefore, the overall time complexity for a single ACO iteration can be given by:
\begin{equation}
\small
\label{chap5-eq-amdvmc-time-complx-1}
\begin{aligned}
T_{\!A\!M\!D\!V\!M\!C_{1}} 	&= T_{2\text{--}7} + T_{11\text{--}21} + T_{23\text{--}29} + T_{31\text{--}36} \\
						&= \mathcal{O}(nAnts.N_{vc}) + \mathcal{O}(nAnts.N_{vc}.N_{pc}) + \mathcal{O}(nAnts) + \mathcal{O}(N_{vc}.N_{pc}) \\
						&= \mathcal{O}(nAnts.N_{vc}.N_{pc}).
\end{aligned}
\end{equation}
And, finally the repeat-until loop of AMDVMC can run for a maximum of $\mathcal{O}(nCycleTerm.nResetMax)$. Therefore, the worst-case time complexity of AMDVMC algorithm can be given by:
\begin{equation}
\small
\label{chap5-eq-amdvmc-time-complx-2}
\begin{aligned}
T_{\!A\!M\!D\!V\!M\!C} 	&= \mathcal{O}(nCycleTerm.nResetMax.nAnts.N_{vc}.N_{pc}).
\end{aligned}
\end{equation}
Furthermore, considering the ACO parameters as constants, the worst-case time complexity can be simplified to the following:
\begin{equation}
\small
\label{chap5-eq-amdvmc-time-complx-3}
\begin{aligned}
T_{\!A\!M\!D\!V\!M\!C} 	&= \mathcal{O}(N_{vc}.N_{pc}).
\end{aligned}
\end{equation}

\paragraph{Memory Overhead:} Similar to MMDVMC, AMDVMC has an $\mathcal{O}(N_{vc}N_{pc})$ memory overhead for maintaining the pheromone information that represents all possible VM-to-PM migration preferences in the cluster and another $\mathcal{O}(nAnts)$ memory overhead for managing $nAnts$ ant agents. In addition, in every iteration, each ant agent generates its migration map using its local list of PMs in the cluster with their associated hosted VMs and updates the VM-to-PM placements. As a result, each ant agent has an additional $\mathcal{O}(N_{vc}N_{pc})$ memory overhead for managing local information of the cluster during each iteration. Therefore, the overall memory overhead of AMDVMC algorithm is the following:
\begin{equation}
\small
\label{chap5-eq-amdvmc-mem-overhead-1}
\begin{aligned}
M_{\!A\!M\!D\!V\!M\!C}	= \mathcal{O}(nAnts.N_{vc}.N_{pc}).
\end{aligned}
\end{equation}
Considering the number of ants is fixed, the memory overhead is simplified as follows:   
\begin{equation}
\small
\label{chap5-eq-amdvmc-mem-overhead-2}
\begin{aligned}
M_{\!A\!M\!D\!V\!M\!C}	= \mathcal{O}(N_{vc}.N_{pc}).
\end{aligned}
\end{equation}

\subsection{Simulation Setup}
\label{chap5-sec-sim-setup}

\subsubsection{Data Center Setup}
\label{chap5-sec-data-center-setup}
The simulated data center consists of $N_{p}$ homogeneous PMs with three-dimensional resource capacities: CPU, memory, and network I/O, and for each PM, the total resource capacities of these resources are set as 5.0 GHz, 10 GB, and 1 Gbps, respectively. Absolute values of PM resource capacities are simulated so that the migration overhead factors can be measured using the proposed migration overhead estimation model. The power consumption for an active PM is calculated according to the power consumption model represented by Eq. 9-10 in our previous work \cite{Ferdaus2014a} and the values for $E_{idle}$ and $E_{full}$ in the model are set to 162 watts and 215 watts, respectively, as used by Gao \textit{et al.} \cite{Gao2013}.

Given the fact that three-tier tree network topology with core-aggregation-access switch levels are predominantly used in production data centers \cite{Kliazovich2013}, PMs in the simulated data center are interconnected with each other using three-tier tree network topology, as shown in Figure \ref{chap5-fig-server-clusters-1}, where each of the network switches have 8 ports. The maximum bandwidth capacity of inter-PM communication links used for the VM migrations is set to 1 Gbps, and the available bandwidths of such links at run-time are synthetically generated using random numbers from the normal distribution with a mean ($Mean\!BW$) of 0.05 and a standard deviation ($S\!D\!B\!W$) of 0.2. The network distance between any two PMs is measured as $D\!S = h \times D\!F$, where $h$ is the number of physical hops (specifically, network switches) between two PMs in the simulated data center architecture as defined above, and $D\!F$ is the \textit{Distance Factor} that implies the physical inter-hop distance. The value of $h$ is computed using the analytical expression for tree topology as presented by Meng et al. \cite{Meng2010}, and $D\!F$ is fed as a parameter to the simulation which is set to 2 for the experiments conducted. The network distance of a PM with itself is set to $0$ which implies that any data communication between two VMs hosted in the same PM is performed by memory copy without going through the physical network. A higher value of $D\!F$ indicates greater relative communication distance between any two data center nodes. 

Before the dynamic consolidation operation, $N_{v}$ number of VMs are considered to be running and are distributed randomly among the $N_{p}$ PMs in a load balanced mode. Such an initial data center state is simulated in order to provide an incentive for the offline consolidation operation so that there is scope for the algorithms to improve resource utilization and reduce energy consumption. The VM resource demands for CPU, memory, and network I/O are synthetically generated using random numbers from the normal distribution with mean $Mean\!Rsc$ and standard deviation $S\!D\!Rsc$. The corresponding VM page dirty rates ($v^{dr}$) are generated using uniform random numbers from the range $[0, PR*v^{mem}]$, where $PR$ is the ratio of maximum possible page dirty rate to VM memory size, and it is set to 0.25 for the whole simulation. Thus, the VM page dirty rate ($v^{dr}$) is effectively parameterized in the simulation by the VM memory demand ($v^{mem}$) and this eliminates the need for another parameter.


Table \ref{chap5-tab-mig-parameters-1} summarizes the values of the parameters used in the various formulations for the VM migration overhead estimation presented in Section \ref{chap5-vm-mig-impact-measurement}. \textcolor{black}{Specifically, the values of the remaining dirty memory threshold ($DV_{th}$), the maximum number of pre-copy migration rounds ($max\_round$), the coefficients for WWS computation ($\mu_{1}$,$\mu_{2}$, and $\mu_{3}$), and the VM resume time ($T_{res}$) are taken as reported from the original paper \cite{Liu2013} and are used in Algorithm \ref{chap5-alg-vmmigoverhead-1}. The coefficients of VM migration energy consumption ($\gamma_{1}$ and $\gamma_{2}$) are used in Eq. \ref{chap5-eq-mec-1} and their values are taken as reported in by Liu \textit{et al.} \cite{Liu2013}.} The coefficients for computing the overall VM migration overhead $M\!O$ ($\alpha_1$, $\alpha_2$, $\alpha_3$, and $\alpha_4$) are used in Eq. \ref{chap5-eq-mo-1} and each of them is set to 0.25 in order to give each of the overhead factors equal weight. The overhead factors are normalized against their maximum possible values before using in formulation \ref{chap5-eq-mo-1}, where the maximum values are identified by conducting preliminary experiments for the above mentioned setup. The percentage of SLA violation ($\sigma$) during any VM migration is used in Eq. \ref{chap5-eq-msv-1} and set to 0.1 as reported in a previous experimental study \cite{Voorsluys2009}. 

Finally, Table \ref{chap5-tab-AMDVMC-parameters-1} shows the optimal values for the input parameters used for the proposed ACO-based, Migration overhead-aware Dynamic VM Consolidation (AMDVMC) algorithm, including those for the ACO metaheuristic. These values are determined by rigorous parameter sensitivity analysis conducted during the preliminary phase of the experiment. Input parameters for other consolidation algorithms are taken as reported in the respective papers. All the experiments presented in this paper have been repeated 1000 times and the average results are reported.

\begin{table}[!t]
\caption{VM migration-related parameters used in the simulation}
\label{chap5-tab-mig-parameters-1}
\centering
\begin{tabular}{|l|l|}
\hline
\multicolumn{1}{|c|}{\textit{Constants/Values}} & \multicolumn{1}{|c|}{\textit{Meaning}}\\
\hline
$DV_{th}=200 (MB)$	&	Remaining dirty memory threshold				\\
\hline
$max\_round=20$		& 	Maximum number of rounds of pre-copy migration 	\\
\hline
$\mu_{1}=-0.0463$	& 	Coefficients for computing Writable Working Set \\
$\mu_{2}=-0.0001$	&	\\
$\mu_{3}=0.3586$	& 	\\
\hline
$T_{res}=20 (ms)$ 	& 	Time needed to resume a VM in the destination PM 	\\
\hline
$\alpha_{1}=0.25$	& 	Coefficients for computing overall VM migration overhead 	\\
$\alpha_{2}=0.25$	&	\\
$\alpha_{3}=0.25$	&	\\
$\alpha_{4}=0.25$	&	\\
\hline
$\gamma_{1}=0.512$	&	Coefficients of VM migration energy computation \\
$\gamma_{2}=20.165$	&	\\
\hline
$\sigma=0.1$		& 	Percentage of SLA violation during migration 	\\
\hline
\end{tabular}
\end{table}


\begin{table}[!t]
\caption{ACS parameter values used for the AMDVMC algorithm in evaluation}
\label{chap5-tab-AMDVMC-parameters-1}
\centering
\begin{tabular}{|c|c|c|c|c|c|c|c|c|c|}
\hline
$nAnts$ & $nCycleTerm$ & $nCycleMax$ & $\beta$ & $\delta$ & $q_{0}$ & $\omega$ & $\lambda$ & $\phi$ \\
\hline
5 & 5 & 100 & 1 & 0.3 & 0.8 & 0.5 & 0.05 & 1 \\
\hline
\end{tabular}
\end{table}

\subsubsection{Performance Evaluation Metrics}
\label{chap5-sec-perf-eval-metrics}
The quality of the consolidation decisions produced by the algorithms are compared across several performance metrics. Each dynamic VM consolidation has two types of performance factors: gain factors and cost factors. 
The gain factors indicate the benefit or profit that can be achieved by a particular consolidation. The first gain factor reported in the results is the number of released PMs in the data center ($nReleasedP\!M$). The consolidation decision that releases the maximum number of PMs effectively consolidates the running VMs in the minimum number of active PMs. The PMs released in this process can be turned to lower power states to reduce power consumption in the data center, or can be utilized to accommodate further VM requests, which effectively improves the capacity of the data center, and eventually, maximizes profit. Another closely related performance metric shown in the results is the packing efficiency ($P\!E$) represented by Eq 12 in our previous work \cite{Ferdaus2014a}. The $P\!E$ indicates the average number of VMs packed or consolidated in each of the active PMs. Therefore, as the term and the formulation suggest, it effectively captures each algorithm's efficiency in packing or consolidating the running VMs in the PMs. 

The overall power consumption (measured in KW) of the active PMs (measured according to Eq. 9-10 in our previous work \cite{Ferdaus2014a}) is reported as the third gain factor in this evaluation. This is one of the most important key performance indicators for any consolidation scheme, which is directly proportional to the operating cost of hosting the running VMs in the data center, since reduction of the power consumption is equivalent to saving on the electricity costs of data center operation. The last factor reported in this category is the overall resource wastage (normalized against the total resource capacity of PM) of the active PMs after the VM consolidation. This factor is measured as the accumulated resource wastage of the PMs that are active after the consolidation where the individual PM's resource wastage (normalized) is measured according to Eq. 8 in our previous work \cite{Ferdaus2014a}. Reduction in resource wastage indicates efficiency in resource utilization in the data center, and thus the consolidation that causes the least amount of resource wastage is to be preferred over others. 
The cost factors reported in this evaluation are the migration overhead factors and associated metrics for achieving a particular dynamic VM consolidation in the data center. As described in earlier sections, dynamic VM consolidation achieved by VM live migrations has adverse effects on the hosted applications and also on the data center. The measures of the cost factors incurred due to a particular consolidation decision, represented by migration map $M\!M$, are captured primarily by the four aggregated migration overhead factors presented in Section \ref{chap5-vm-mig-impact-measurement}: the estimate of aggregated data (memory) to be transmitted across the data center due to the VM migrations $M\!D(\!M\!M)$ (Eq. \ref{chap5-eq-md-1}) in terabytes (TB), the aggregated migration time $MT(\!M\!M)$ (Eq. \ref{chap5-eq-mt-1}) and the aggregated VM downtime $DT(\!M\!M)$ (Eq. \ref{chap5-eq-dt-1}), both in the number of hours, and the aggregated network cost $N\!C(\!M\!M)$ (Eq. \ref{chap5-eq-nc-1}). Obviously, for all of these cost factors, the VM consolidation decision that results in the lowest overhead factors will be preferable over others. Moreover, the unified migration overhead $M\!O(\!M\!M)$ (Eq. \ref{chap5-eq-mo-2}) is also reported as a single metric that captures the overall migration impact of a consolidation. Furthermore, an estimate of the aggregated migration energy consumption $M\!EC(\!M\!M)$ (in Kilo Joules) by the data center components (Eq. \ref{chap5-eq-mec-2}) and an estimate of the aggregated SLA violation $M\!SV(\!M\!M)$ (Eq. \ref{chap5-eq-msv-2}) of hosted applications due to VM migrations are reported. 

All the above performance metrics are measured against the following scaling factors: (1) DC size ($N_{p}$), (2) mean resource demands of VMs ($MeanRsc$), and (3) diversification of workloads ($SDRsc$). The following subsections present the results and analysis for each of the experiments conducted.

\subsubsection{Simulation Environment} 
The offline simulation environment and the compared algorithms are implemented in Java (JDK and JRE version 1.7.0) and the simulation is conducted on a Dell Workstation (Intel Core i5-2400 3.10 GHz CPU (4 cores), 4 GB of RAM, and 240 GB storage) hosting Windows 7 Professional Edition.

\subsection{Scaling Data Center Size}
\label{chap5-sec-scale-dc-size}
This part of the experiment demonstrates the quality of the VM consolidation decisions produced by the algorithms with increasing problem size--- the number of PMs ($N_{p}$) in the data center is set to 64 and increased up to 4096 in stages, each step doubling the previous number. The number of VMs running in the data center is derived from the simulated number of PMs: $N_{v} = 2*N_{p}$. As for the other parameters, $MeanRsc$ and $SDRsc$ are set to 0.05 and 0.2, respectively. 

\begin{figure}[!t]
\centering
\includegraphics[scale=0.45, trim=.5cm 9.5cm 0cm 1.5cm]{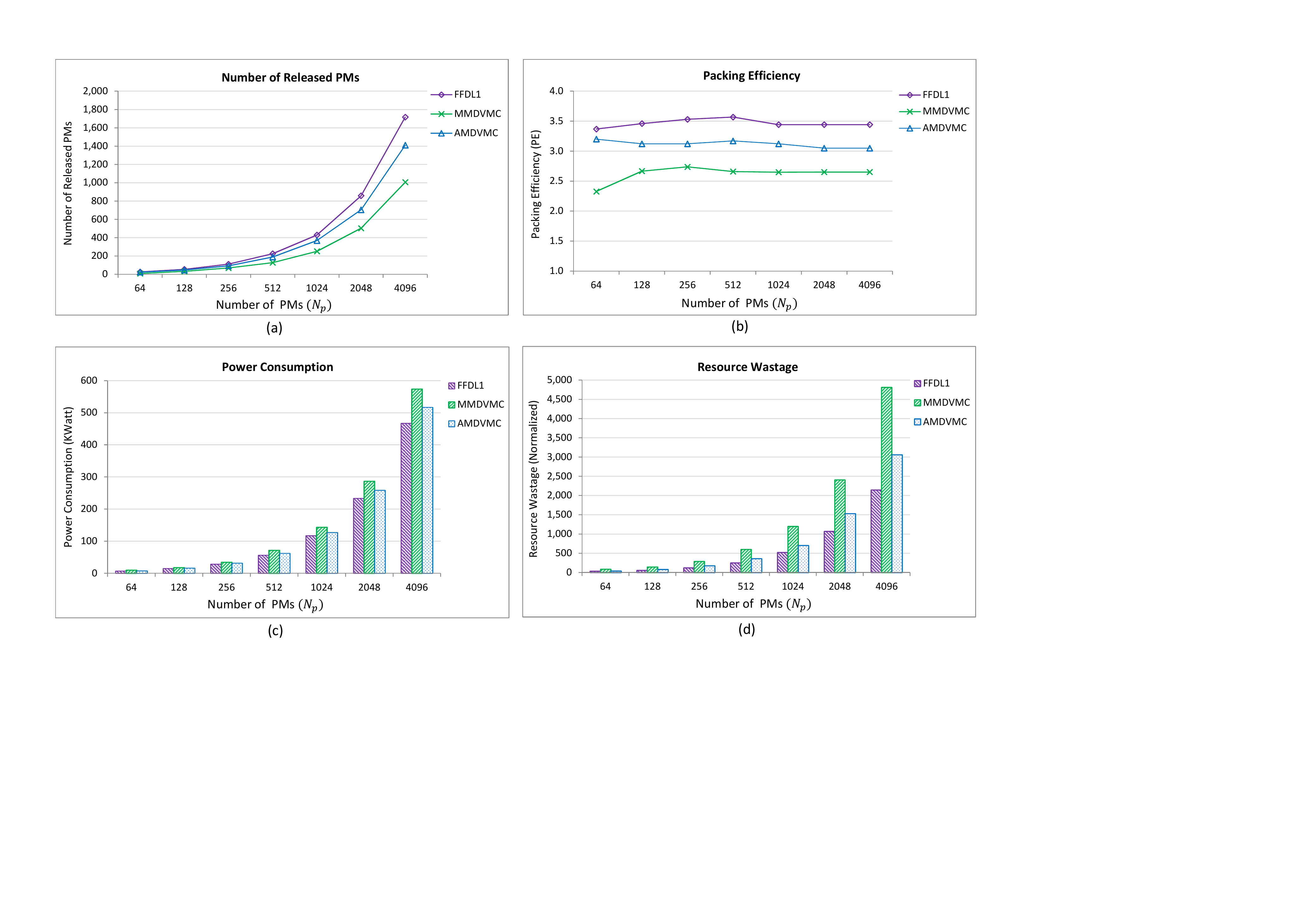}
\caption{Performance of the algorithms with increasing $N_{p}$: (a) Number of Released PMs, (b) Packing Efficiency, (c) Power Consumption, and (d) Resource Wastage.}
\label{chap5-fig-perf-nPM-1}
\end{figure}

Figure \ref{chap5-fig-perf-nPM-1} shows the four gain factors mentioned above that resulted from the VM consolidation decisions produced by the algorithms for different data center sizes. The average number of PMs released by the algorithms for each $N_{p}$ value is plotted in Figure \ref{chap5-fig-perf-nPM-1}(a). As the figure demonstrates, on average, FFDL1, MMDVMC, and AMDVMC algorithms released 42\%, 23\%, and 36\% of the PMs, respectively, for different data center sizes. FFDL1, being migration-unaware, consolidates the VMs without any regard to the current VM-to-PM placements and therefore, released the maximum number of PMs. MMDVMC, on the other hand, released the least number of PMs, given that it tried to keep the number of VM migrations minimal at the same time. Finally, the performance of the proposed AMDVMC algorithm lies between the other two algorithms by releasing 15\% fewer PMs compared to FFDL1 and 63\% more PMs than MMDVMC. This is also reflected in Figure \ref{chap5-fig-perf-nPM-1}(b) where it is observed that AMDVMC achieved an average PE of 3.1, whereas FFDL1 and MMDVMC achieved PEs of 3.5 and 2.6, respectively. 

Similar performance patterns are demonstrated in Figure \ref{chap5-fig-perf-nPM-1}(c) and Figure \ref{chap5-fig-perf-nPM-1}(d) which show the average power consumption and the normalized resource wastage of the active PMs after consolidation, respectively. It can be seen from the figures that both the power consumption and the resource wastage increase at the same rates as the number of PMs ($N_{p}$) are increased in the data center. Furthermore, compared to MMDVMC, the consolidation decisions produced by AMDVMC result in 13\% less power consumption on average and 42\% less resource wastage, respectively, whereas compared to FFDL1, AMDVMC incurs 9\% more average power consumption and 38\% more resource wastage. Therefore, it is evident from these results that AMDVMC performs better in terms of power consumption and resource wastage compared to the other migration-aware approach, whereas the migration-unaware approach beats AMDVMC in these metrics.

\begin{figure}[!t]
\centering
\includegraphics[scale=0.45, trim=-1cm 10cm 0cm 1.5cm]{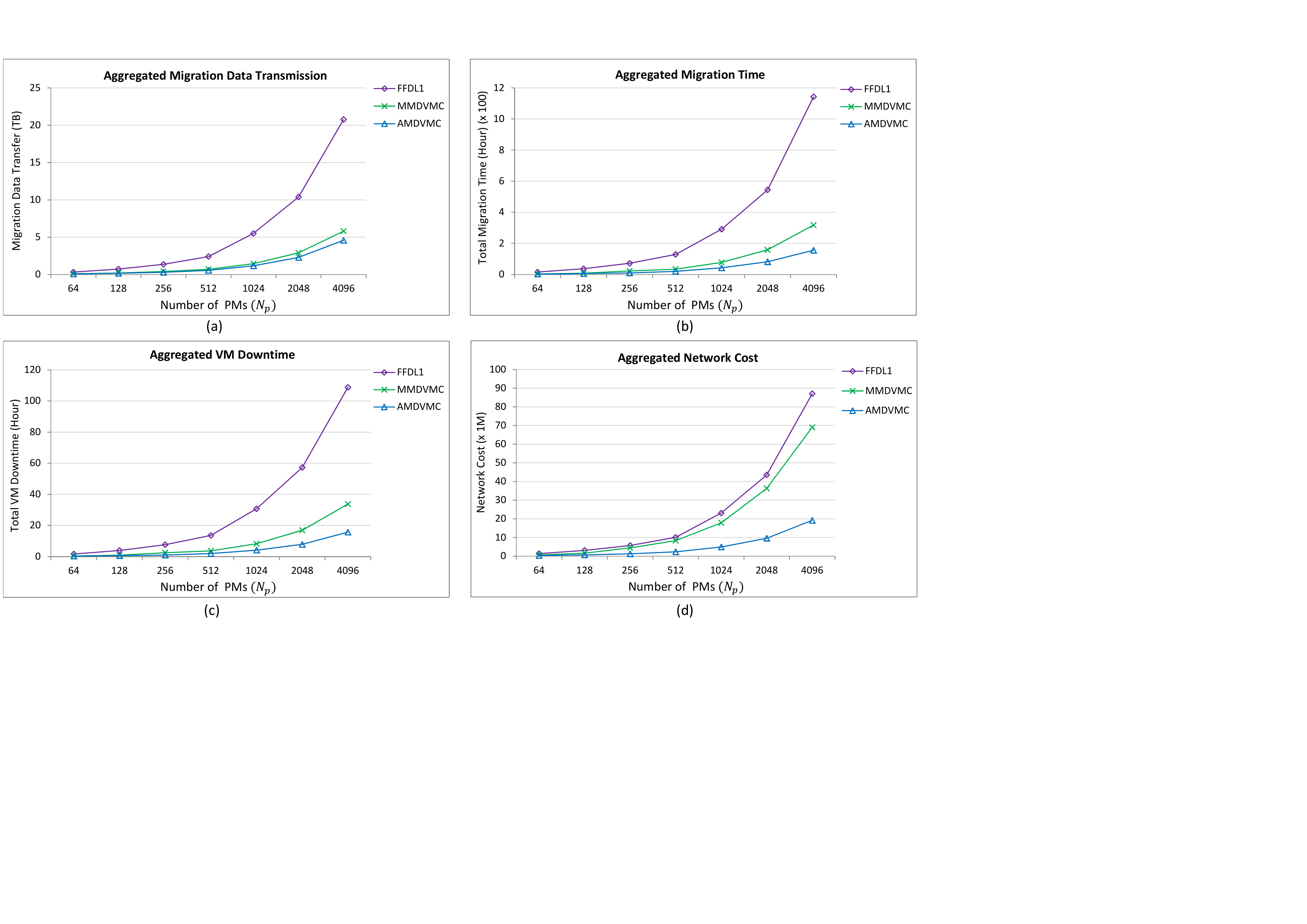}
\caption{Performance of the algorithms with increasing $N_{p}$: (a) Aggregated Migration Data Transmission, (b) Aggregated Migration Time, (c) Aggregated VM Downtime, and (d) Aggregated Network Cost.}
\label{chap5-fig-perf-nPM-2}
\end{figure}

Figure \ref{chap5-fig-perf-nPM-2} shows the four primary cost factors of dynamic consolidation decisions produced by the algorithms for various data center sizes. The estimate of the aggregated amount of VM memory data to be transmitted across the data center due to VM migrations is plotted in Figure \ref{chap5-fig-perf-nPM-2}(a). As the figure depicts, the data transmission rises sharply for FFDL1 with the increasing number of PMs. This is due to the fact that FFDL1 is migration-unaware and therefore, causes many VM migrations which result in a large amount of VM memory data transmission. MMDVMC, being multi-objective, tries to reduce the number of migrations and therefore, causes a lower amount of migration related data transmission. Lastly, AMDVMC is also a multi-objective consolidation approach which takes the estimate of memory data transfer into account during the solution-building process and as a consequence, it incurs the least amount of data transmission relating to VM migrations. In summary, on average, AMDVMC resulted in 77\% and 20\% less migration data transmission compared to FFDL1 and MMDVMC, respectively. For the aggregated migration time and VM downtime, a similar performance pattern can be observed from Figure \ref{chap5-fig-perf-nPM-2}(b) and Figure \ref{chap5-fig-perf-nPM-2}(c), respectively, where both the values increase at a proportional rate with the increase of $N_{p}$. This is reasonable since the number of VMs ($N_{v}$) increases in proportion to the number of PMs ($N_{p}$), which in turn contributes to the proportional rise of aggregated migration time and VM downtime. Compared to FFDL1 and MMDVMC, on average, AMDVMC caused 84\% and 85\% less aggregated migration time, and 85\% and 43\% less aggregated VM downtime across all data center sizes.
Figure \ref{chap5-fig-perf-nPM-2}(d) shows the estimate of aggregated network cost due the VM migrations for the consolidation decisions. The figure shows that for both FFDL1 and MMDVMC, the network cost increases sharply with the number of PMs in the data centers, whereas it increases slowly for AMDVMC. This is due to the fact that FFDL1 is migration overhead-unaware and MMDVMC, although it is in a way migration-aware, forms neighborhoods of PMs randomly for performing consolidation operations and therefore, does not take any type of network cost into account while making migration decisions. On average, the observed network cost improvements of AMDVMC over FFDL1 and MMDVMC are 77\% and 65\%, respectively.

\begin{figure}[!t]
\centering
\includegraphics[scale=0.45, trim=-1cm 11.5cm 0.2cm .6cm]{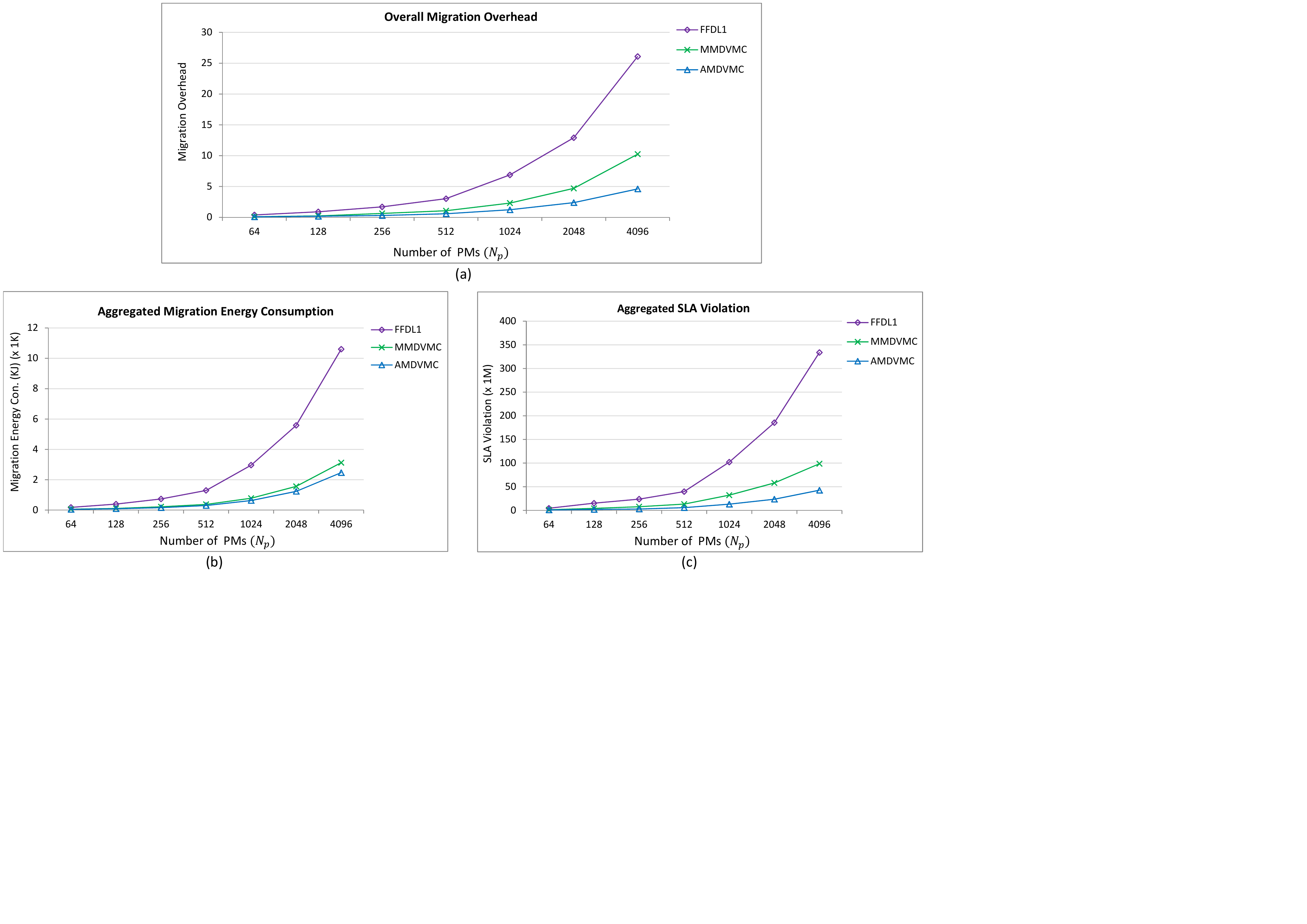}
\caption{Performance of the algorithms with increasing $N_{p}$: (a) Overall Migration Overhead, (b) Aggregated Migration Energy Consumption, and (c) Aggregated SLA Violation.}
\label{chap5-fig-perf-nPM-3}
\end{figure}

Figure \ref{chap5-fig-perf-nPM-3}(a) presents a summary of the overall migration overhead incurred by the algorithms as per formulation \ref{chap5-eq-mo-2} where, on average, AMDVMC incurs 81\% and 38\% less migration overhead compared to FFDL1 and MMDVMC, respectively. Furthermore, the estimate of aggregated migration energy consumption and SLA violation are shown in Figure \ref{chap5-fig-perf-nPM-3}(b) and Figure \ref{chap5-fig-perf-nPM-3}(c), respectively. Since such energy consumption and SLA violation depend on the migration-related data transmission and migration time, respectively, these figures have similar performance patterns as those of Figure \ref{chap5-fig-perf-nPM-1}(a) and Figure \ref{chap5-fig-perf-nPM-1}(b), respectively. In summary, compared to FFDL1 and MMDVMC, on average AMDVMC reduces the migration energy consumption by 77\% and 20\%, and SLA violation by 85\% and 52\%, respectively.

From the results and discussions presented above, it can be concluded that, for all three compared algorithms, both the gain factors and the cost factors increase at a proportional rate with the size of the data center ($N_{p}$). In comparison to the migration-aware MMDMVC approach, the proposed AMDVMC scheme outperforms MMDVMC on both gain factors and cost factors by generating more efficient VM consolidation plans that result in reduced power consumption, resource wastage, and migration overhead. On the other hand, FFDL1, being migration-unaware, generates VM consolidation plans that result in lower power consumption and resource wastage compared to AMDVMC; however, this is achieved at the cost of much higher migration overhead factors.

\subsection{Scaling Mean Resource Demand}
\label{chap5-sec-scale-mean-rsc-dem}
In order to compare the quality of the solutions produced by the algorithms for various sizes of the active VMs, this part of the experiment starts with a mean VM resource demand ($M\!eanRsc$) of 0.05 and increases it up to 0.3, raising it each time by 0.05. The maximum value for $M\!eanRsc$ is kept at 0.3 in order to ensure that the VMs are not too large compared to the PM so that there will be little scope for performing consolidation operations. Moreover, multi-dimensionality of resource types reduces the scope of VM consolidation. Otherwise, if on average, only one VM can be assigned per PM, there is no way of consolidating VMs and releasing PMs to improve power and resource efficiency. The number of PMs ($N_{p}$) in the simulated data center is set at 1024 and the number of simulated active VMs ($N_{v}$) in the data center is derived from the number of PMs and mean VM resource demands using the following formulation: 
\begin{equation}
\small
\label{chap5-eq-eval-Nv-MeanRsc-1}
N_{v} = N_{p}*(0.55-M\!eanRsc)/0.25.
\end{equation}
Table \ref{chap5-tab-eval-Nv-MeanRsc-1} shows the different values for $N_{v}$ produced by the above equation for each $M\!eanRsc$ value. This approach ensures that for the initial states, on average, each PM hosts two VMs when $M\!eanRsc=0.05$ and with a gradual increase of $M\!eanRsc$, the average number of VMs hosted by each PM is reduced up to a point where, when $M\!eanRsc=0.30$, each PM hosts one VM in the initial state. Such an approach creates initial states that offer scope for VM consolidation and therefore, provides opportunities for comparing the efficiency of VM consolidation algorithms. For all these cases, the standard deviation of VM resource demand $S\!D\!Rsc$ is set to 0.2. 

\begin{table}
\parbox{.45\linewidth}{
\centering \caption{Number of VMs ($N_{v}$) for corresponding $M\!eanRsc$ values} \label{chap5-tab-eval-Nv-MeanRsc-1}
	\begin{tabular}{|c|c|c|}
		\hline
		$M\!eanRsc$	& $N_{p}$ 	& $N_{v}$ \\
		\hline
		0.05	& 1024	& 2048	\\
		0.10	& 1024	& 1843	\\
		0.15	& 1024	& 1638	\\
		0.20	& 1024	& 1434	\\
		0.25	& 1024	& 1229	\\
		0.30	& 1024	& 1024 	\\
		\hline
	\end{tabular}
}
\hfill
\parbox{.45\linewidth}{
\centering \caption{Number of VMs ($N_{v}$) for corresponding $S\!D\!Rsc$ values} \label{chap5-tab-eval-Nv-SDRsc-1}
	\begin{tabular}{|c|c|c|}
		\hline
		$S\!D\!Rsc$	& $N_{p}$ 	& $N_{v}$ \\
		\hline
		0.05	& 1024	& 2048	\\
		0.10	& 1024	& 1843	\\
		0.15	& 1024	& 1638	\\
		0.20	& 1024	& 1434	\\
		0.25	& 1024	& 1229	\\
		0.30	& 1024	& 1024 	\\
		\hline
	\end{tabular}
}
\end{table}

\begin{figure}[!t]
\centering
\includegraphics[scale=0.45, trim=.5cm 10cm 0cm 1.5cm]{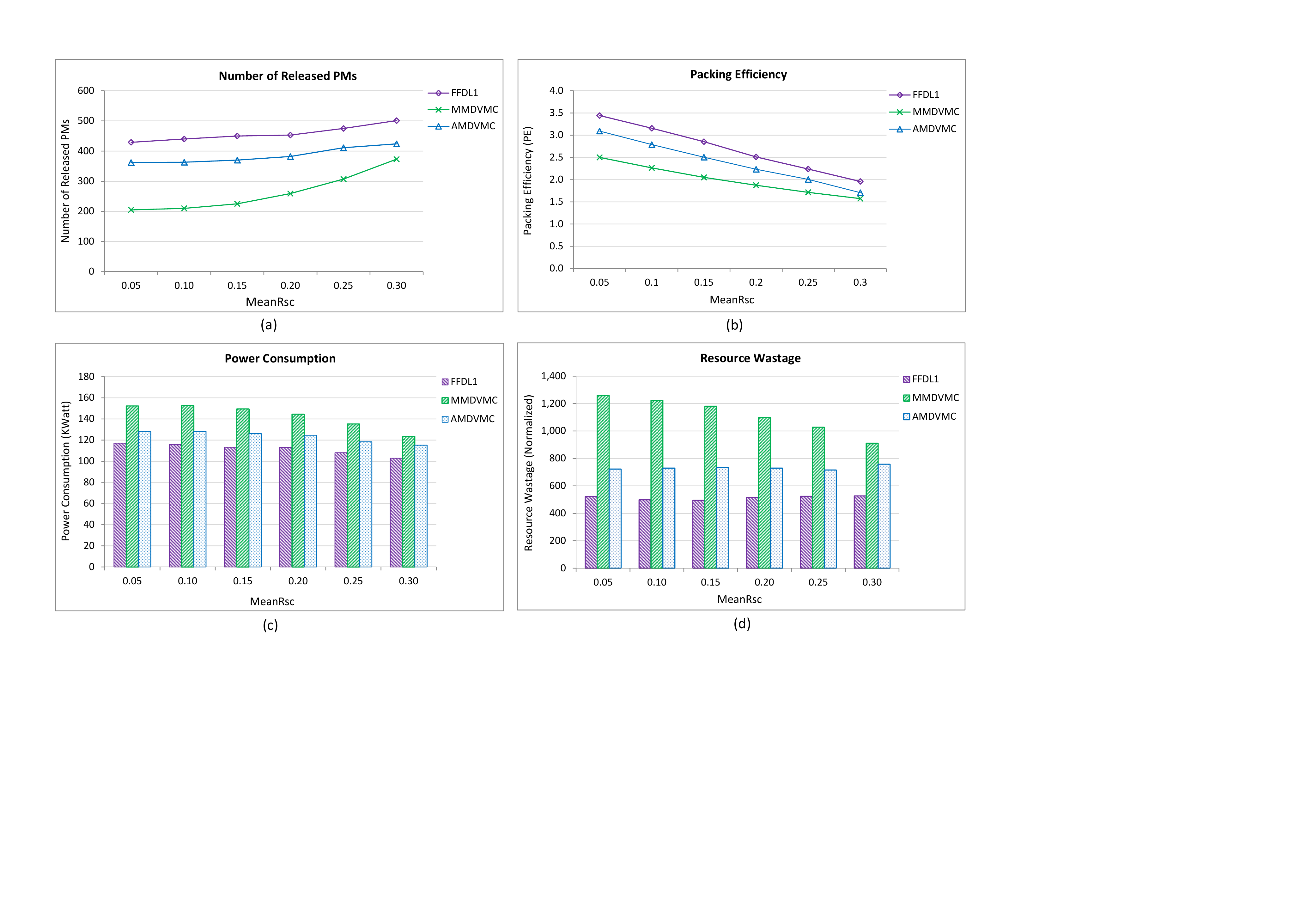}
\caption{Performance of the algorithms with increasing $M\!eanRsc$: (a) Number of Released PMs, (b) Packing Efficiency, (c) Power Consumption, and (d) Resource Wastage.}
\label{chap5-fig-perf-MeanRscSame-1}
\end{figure}

The four gain factors for each of the algorithms for the various means of VM resource demand are plotted in Figure \ref{chap5-fig-perf-MeanRscSame-1}. It can be observed from Figure \ref{chap5-fig-perf-MeanRscSame-1}(a) that the number of PMs released by each of the algorithms gradually increases as the $M\!eanRsc$ increases. This is due to the fact that the number of VMs in the data center decreases with the increase of $M\!eanRsc$ and, as a result, more PMs are released by the algorithms even though the VM size increases. On average, FFDL1, MMDVMC, and AMDVMC have released 45\%, 26\%, and 38\% of PMs in the data center, respectively. In contrast to Figure \ref{chap5-fig-perf-MeanRscSame-1}(a), the packing efficiency $PE$ for each of the algorithms decreases consistently with the increase of $M\!eanRsc$ (Figure \ref{chap5-fig-perf-MeanRscSame-1}(b)). This makes sense since PM's packing efficiency is reduced when packing larger VMs. On average, FFDL1, MMDVMC, and AMDVMC achieve PEs of 2.7, 2.0, and 2.4, respectively. Furthermore, for all the algorithms, the power consumption of the active PMs is reduced with the increase of $M\!eanRsc$, as depicted by Figure \ref{chap5-fig-perf-MeanRscSame-1}(c). With the increase of mean VM resource demands, the algorithms release more PMs and that indicates that the VMs are packed into a reduced number of active PMs, which causes a reduction in power consumption. Compared to MMDVMC, on average, AMDVMC reduces the power consumption by 13\%, whereas it incurs 11\% more power consumption than to FFDL1. Figure \ref{chap5-fig-perf-MeanRscSame-1}(d) shows the resource wastage of the active PMs in the data center. On average, compared to MMDVMC, AMDVMC reduces resource wastage by 34\%, whereas it incurs 42\% more resource wastage compared to FFDL1. Moreover, with the increase of $M\!eanRsc$, resource wastage is reduced gradually for MMDVMC, which indicates that MMDVMC utilizes multi-dimensional resources better for larger VMs compared to smaller VMs. However, in the case of FFDL1 and AMDVMC, the resource wastage gradually reduces for smaller VM sizes. Therefore, it can be concluded from the results that, similar to the results for scaling $N_{p}$, for the gain factors, the AMDVMC algorithm outperforms the migration-aware MMDVMC algorithm, while AMDVMC performs poorly compared to the migration-unaware FFDL1 algorithm.

\begin{figure}[!t]
\centering
\includegraphics[scale=0.45, trim=-.6cm 10cm 0cm 1.8cm]{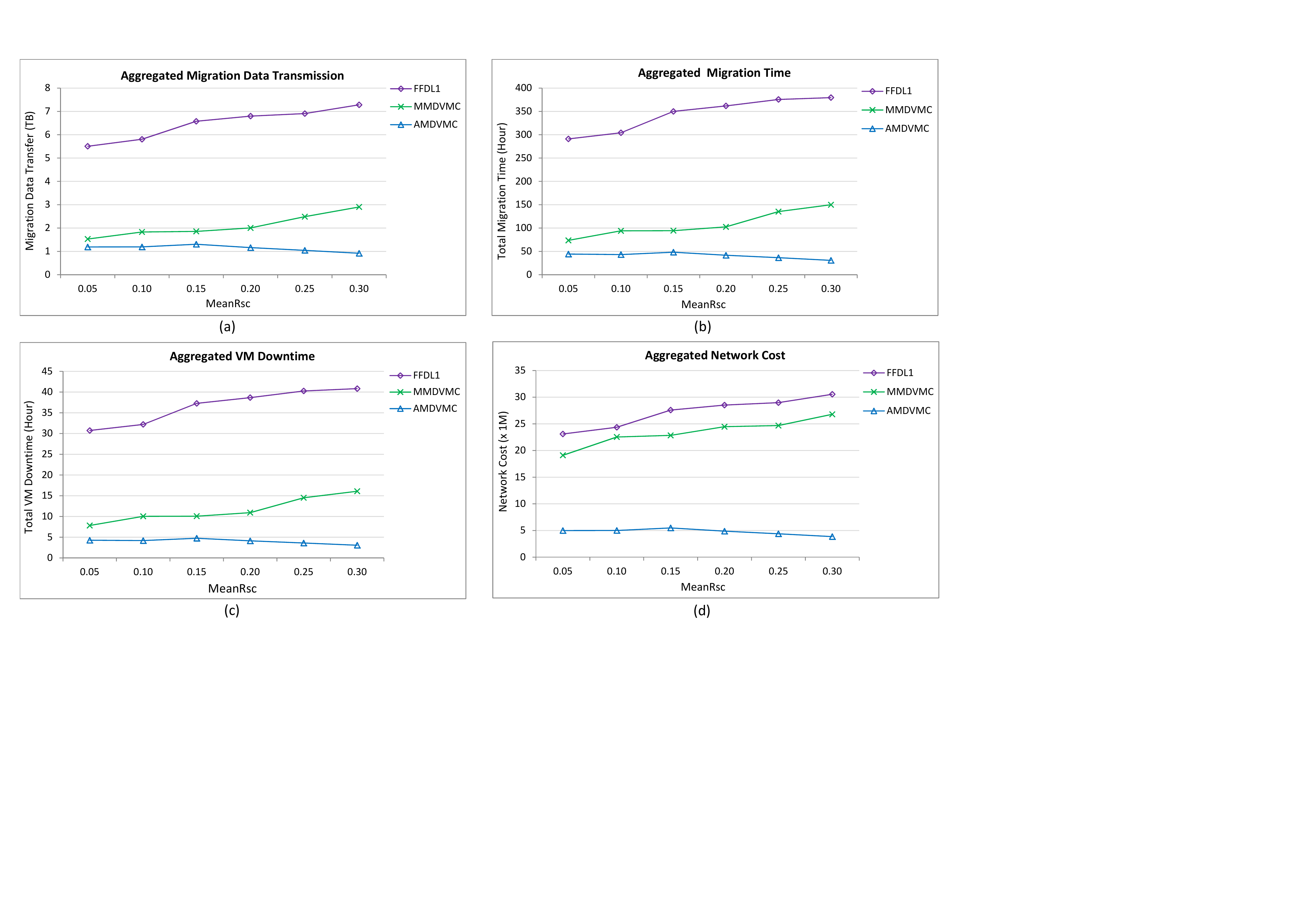}
\caption{Performance of the algorithms with increasing $M\!eanRsc$: (a) Aggregated Migration Data Transmission, (b) Aggregated Migration Time, (c) Aggregated VM Downtime, and (d) Aggregated Network Cost.}
\label{chap5-fig-perf-MeanRscSame-2}
\end{figure}

Figure \ref{chap5-fig-perf-MeanRscSame-2} presents the four primary cost factors of dynamic VM consolidation decisions generated by the algorithms for different means of VM resource demands. Figure \ref{chap5-fig-perf-MeanRscSame-2}(a) shows how the estimate of aggregated migration data transmission changes with respect to $M\!eanRsc$. FFDL1, being migration-unaware, requires an increasing amount of migration-related data transmission as $M\!eanRsc$ increases. This is because, with the increase of $M\!eanRsc$, memory sizes of the VMs also increase, which in turn contributes to the rise in migration data transfer (Algorithm \ref{chap5-alg-vmmigoverhead-1}). MMDVMC, on the other hand, although aims at minimizing the number of migrations, it does not consider the VM memory sizes while making migration decisions and thereby, assumes every VM migration has the same migration overhead, and as consequence, its migration data transfer also increases with the increase of VM sizes. Lastly, in the case of AMDVMC, the estimate of migration data transfer is reduced with the increase of $M\!eanRsc$. This is because AMDVMC considers the estimate of migration data transfer as a contributory factor for the migration overhead estimation; thus it takes this overhead factor into account while making VM consolidation decisions. As a result, with the increase of $M\!eanRsc$ (consequently, VM memory sizes), AMDVMC makes efficient selections of VMs for migration that in turn reduces the migration data transfer. On average, AMDVMC incurs 82\% and 43\% less migration data transfer compared to FFDL1 and MMDVMC, respectively.

Similar performance traits can be observed from Figure \ref{chap5-fig-perf-MeanRscSame-2}(b) and Figure \ref{chap5-fig-perf-MeanRscSame-2}(c) that show the estimates of aggregated migration time and VM downtime. With the increase of $M\!eanRsc$, both the migration time and VM downtime increase for FFDL1 and MMDVMC, whereas these values decrease for AMDVMC. This is due to the same reason as explained for migration data transmission metric. On average, compared to FFDL1 and MMDVMC, AMDVMC reduces the aggregated migration time by 88\% and 59\%, and the aggregated VM downtime by 89\% and 63\%, respectively. It can be further observed from Figure \ref{chap5-fig-perf-MeanRscSame-2}(d) that, with the increase of $M\!eanRsc$, the estimate of the aggregated network cost for FFDL1 and MMDVMC increase gradually, whereas it decreases for AMDVMC. Since with the increase of $M\!eanRsc$, VM memory sizes increase and the network cost is proportional to the amount of migration data transmission, both FFDL1 and MMDVMC incurs much higher network cost compared to that of AMDVMC which is aware of the network cost. On average, AMDVMC shows 82\% and 79\% improvements in this metric compared to FFDL1 and MMDVMC, respectively.

\begin{figure}[!t]
\centering
\includegraphics[scale=0.45, trim=-1cm 11.5cm 0.2cm .6cm]{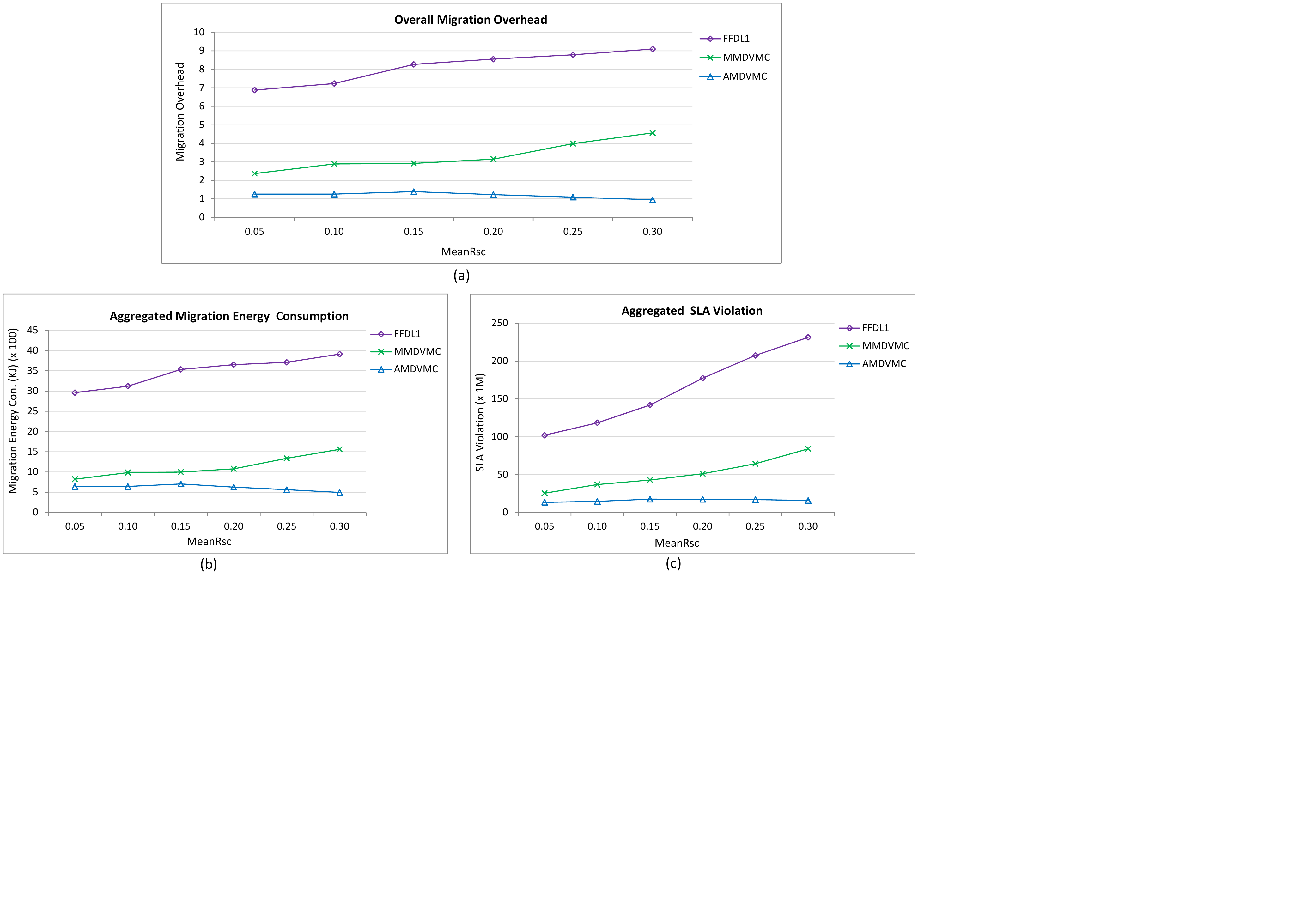}
\caption{Performance of the algorithms with increasing $M\!eanRsc$: (a) Overall Migration Overhead, (b) Aggregated Migration Energy Consumption, (c) Aggregated SLA Violation.}
\label{chap5-fig-perf-MeanRscSame-3}
\end{figure}

For various $M\!eanRsc$ values, the overall migration overhead (Eq. \ref{chap5-eq-mo-2}) is depicted in Figure \ref{chap5-fig-perf-MeanRscSame-3}(a) which shows that AMDVMC incurs 85\% and 61\% less migration overhead compared to FFDL1 and MMDVMC, respectively. Figure \ref{chap5-fig-perf-MeanRscSame-3}(b) and Figure \ref{chap5-fig-perf-MeanRscSame-3}(c) present the estimate of the aggregated migration energy consumption and SLA violation due to the VM migrations for various VM sizes. Since both FFDL1 and MMDVMC do not take into account the migration overhead factors while making consolidation decisions, both the metrics increase with respect to VM memory size. In summary, compared to FFDL1 and MMDVMC, AMDVMC reduces the aggregated migration energy consumption by 82\% and 42\%, and SLA violation by 89\% and 64\%, respectively.

In light of the above results and discussion, it can be summarized that, with the gradual increase of mean VM resource demand, both the power consumption and resource wastage of the data center slowly reduces for both FFDL1 and MMDVMC, whereas for AMDVMC the power consumption reduces slowly, but the resource wastage slightly increases. However, with the increase of $M\!eanRsc$, the cost factors steadily increase for both FFDL1 and MMDMVC, whereas they remain almost steady for AMDVMC. When compared with the migration-aware MMDMVC approach, the proposed AMDVMC algorithm outpaces MMDVMC on both the gain and cost factors, thereby indicating the superior quality of the VM consolidation plans produced by AMDVMC. In contrast, the FFDL1 algorithm produces VM consolidation plans that require less power consumption and resource wastage compared to AMDVMC; however, this migration-unaware approach results in much higher migration overhead.

\subsection{Diversification of Workload}
\label{chap5-sec-div-workload}
This part of the experiment was conducted to assess the algorithms by diversifying the workloads of the VMs. This was done by varying the standard deviation of the VM resource demands ($S\!D\!Rsc$), where the initial value of $S\!D\!Rsc$ is set to 0.05 and it is gradually increased up to 0.3, with an increment of 0.05 each time. Similar to the approach applied for scaling $M\!eanRsc$, the maximum value of $S\!D\!Rsc$ was kept at 0.3 so that the VM's resource demand for any resource dimension (e.g., CPU, memory, or network I/O) was not too large compared to the PM's resource capacity for that corresponding resource dimension and by this way, it helps to keep scope of consolidation. Similar to formulation \ref{chap5-eq-eval-Nv-MeanRsc-1}, the number of VM was derived using the following, while keeping $N_{p} = 1024$:
\begin{equation}
\small
\label{chap5-eq-eval-Nv-SDRsc-1}
N_{v} = N_{p}*(0.55-SDRsc)/0.25.
\end{equation}
Table \ref{chap5-tab-eval-Nv-SDRsc-1} shows the different values for $N_{v}$ produced by the above equation for each $S\!D\!Rsc$ value. And, the mean VM resource demand $M\!eanRsc$ was set to 0.05.

\begin{figure}[!t]
\centering
\includegraphics[scale=0.45, trim=.5cm 9.5cm 0cm 1.5cm]{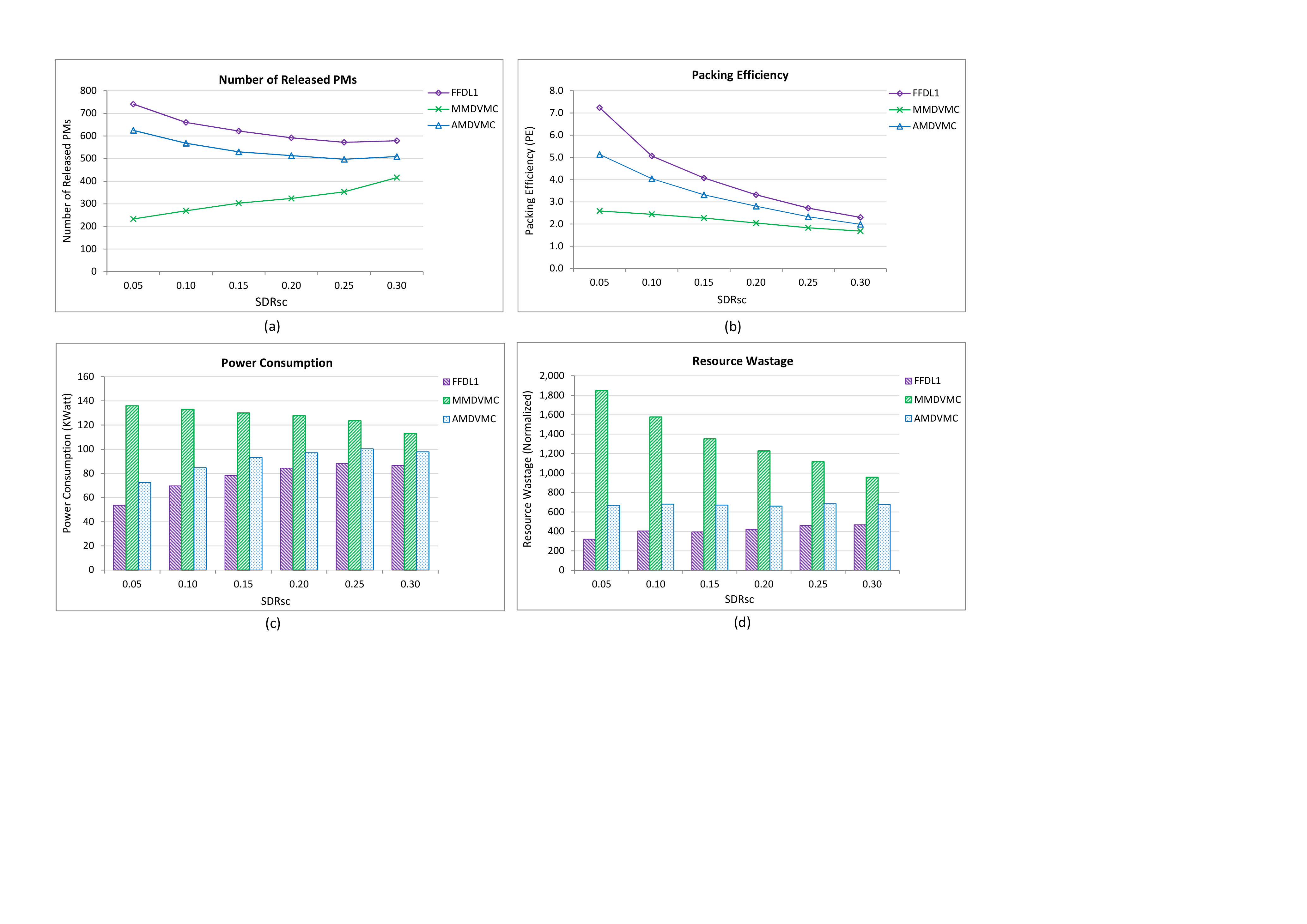}
\caption{Performance of the algorithms with increasing $S\!D\!Rsc$: (a) Number of Released PMs, (b) Packing Efficiency, (c) Power Consumption, and (d) Resource Wastage.}
\label{chap5-fig-perf-SDRsc-1}
\end{figure}

Figure \ref{chap5-fig-perf-SDRsc-1} presents the four gain factors for the algorithms while scaling the standard deviation $S\!D\!Rsc$ of VM resource demands. It can be observed from Figure \ref{chap5-fig-perf-SDRsc-1}(a) that, with the increase of workload diversification, the number of PMs released gradually decreases for FFDL1 and AMDVMC, whereas the opposite trend is found for MMDVMC. This can be explained as follows. Since FFDL1 works with the greedy strategy of First Fit, when the variation in the amount of resource demands for different resource types increases, placement feasibility for the VMs decreases, and as a consequence, FFDL1 requires relatively more active PMs for higher $S\!D\!Rsc$ values. However, MMDVMC utilizes the MMAS metaheuristic \cite{Stutzle2000} which is an iterative solution refinement process and therefore, can be effective even though resource demand variation is high. And, even though AMDVMC utilizes the ACO metaheuristic \cite{Dorigo1997}, being multi-objective, it also targets in reducing the migration overhead and as a result, its performance in terms of gain factors reduces with the increase of $S\!D\!Rsc$, which effectively increases the VM memory size for a portion of the VMs in the data center. Nevertheless, when compared among the algorithms, the proposed AMDVMC outperforms MMDVMC by releasing 79\% more PMs on average, whereas it release 14\% fewer PMs compared to the migration-unaware FFDL1.

With the increase of $SDRsc$, the probability of generating VMs with higher resource demands across the resource dimensions also increases and thereby, the packing efficiency of the algorithms gradually decreases, which is reflected in Figure \ref{chap5-fig-perf-SDRsc-1}(b). On average, FFDL1, MMDVMC, and AMDVMC achieve PEs of 4.1, 2.1, and 3.3, respectively. Figure \ref{chap5-fig-perf-SDRsc-1}(c) and Figure \ref{chap5-fig-perf-SDRsc-1}(d) demonstrate similar performance trend as observed in Figure \ref{chap5-fig-perf-SDRsc-1}(a). For FFDL1 and AMDVMC, since the number of active PMs increases with the respect to $SDRsc$, both the power consumption and resource wastage of active PMs gradually increase, whereas these metrics are reduced for MMDVMC. Finally, on average, compared to MMDVMC, AMDVMC reduces the power consumption and resource wastage by 28\% and 48\% respectively, whereas, compared to FFDL1, it incurs 20\% more power consumption and 66\% more resource wastage. Therefore, in the context of the gain factors, it can be concluded from the above results that similar to results for scaling $N_{p}$ and $M\!eanRsc$, AMDVMC outperforms the migration-aware MMDVMC algorithm, while the migration-unaware FFDL1 algorithm performs better than AMDVMC. 

\begin{figure}[!t]
\centering
\includegraphics[scale=0.45, trim=-1cm 11.6cm 3cm 0cm]{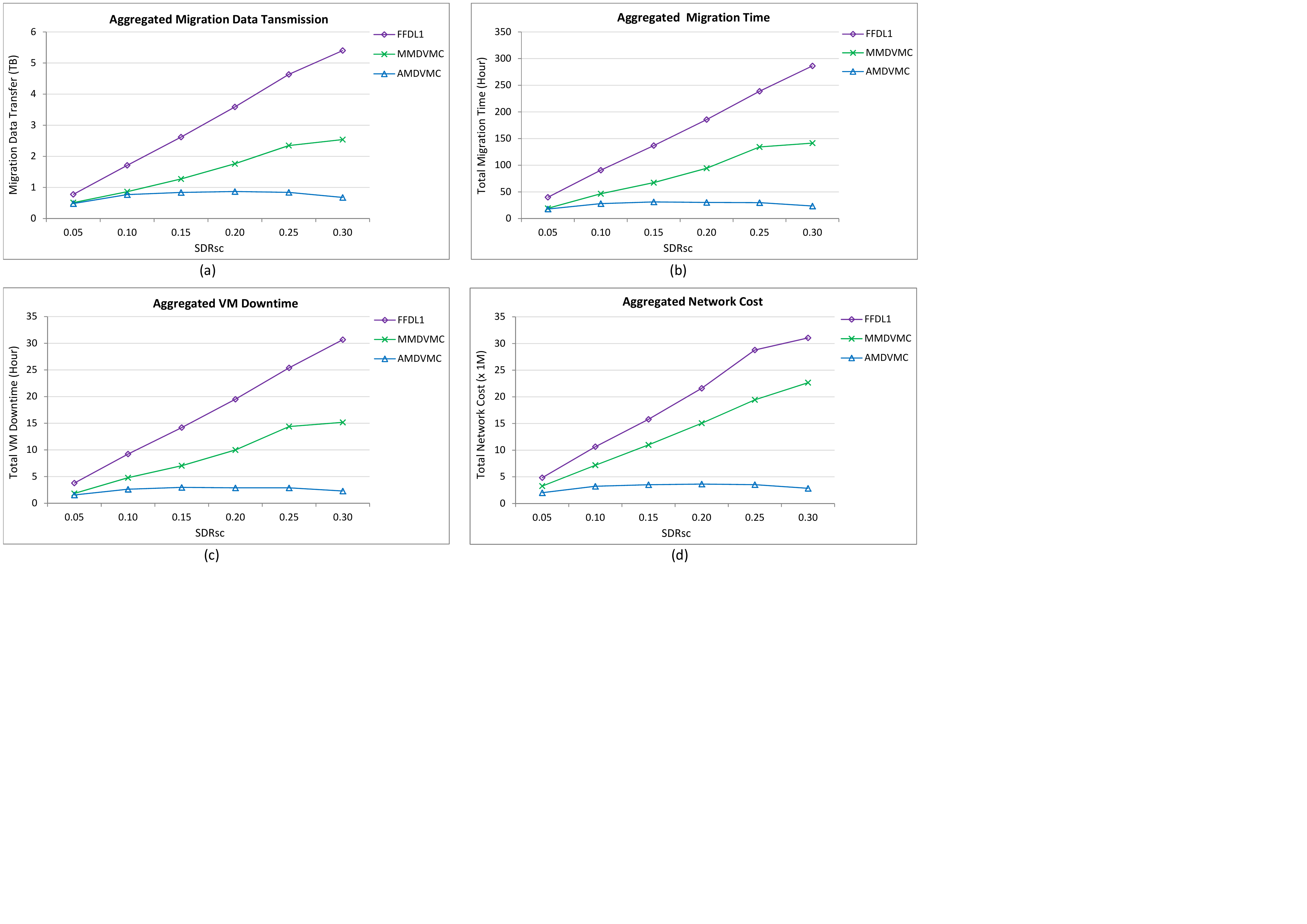}
\caption{Performance of the algorithms with increasing $S\!D\!Rsc$: (a) Aggregated Migration Data Transmission, (b) Aggregated Migration Time, (c) Aggregated VM Downtime, and (d) Aggregated Network Cost.}
\label{chap5-fig-perf-SDRsc-2}
\end{figure}

Figure \ref{chap5-fig-perf-SDRsc-2} shows the four primary cost factors with increasing diversity of workloads ($SDRsc$). As depicted in Figure \ref{chap5-fig-perf-SDRsc-2}(a), the estimates of the aggregated migration data transmission for both FFDL1 and MMDVMC increase with respect to VM workload variation. With the increase of $SDRsc$, more VMs tend to have larger memory sizes and as a consequence, migration data transmission for the migration-unaware FFDL1 algorithm increases steadily. In the case of MMDVMC, it is worth noting that it improves the gain factors steadily with the increase of $SDRsc$ (Figure \ref{chap5-fig-perf-SDRsc-1}), which is achieved at the cost of steady increase in migration data transmission and other cost factors (Figure \ref{chap5-fig-perf-SDRsc-2}). And, for AMDVMC, the migration data transmission slightly increases upto $SDRsc = 0.2$ and thereafter, it decreases. The increase for the cases when $SDRsc \leq 0.2$ is explained by the fact that, as $SDRsc$ increases, the VM resource demands (including VM memory size) increases probabilistically, which in turn raises the migration data transmission. However, with the increase of $SDRsc$, the number of VMs ($N_{v}$) decreases according to Eq. \ref{chap5-eq-eval-Nv-SDRsc-1} and AMDVMC, being migration overhead-aware, can reduce the migration data transmission better for a relatively smaller number of VMs when $SDRsc > 0.2$, compared to the cases when $SDRsc \leq 0.2$. On average, compared to FFDL1 and MMDVMC, AMDVMC requires 68\% and 40\% less migration data transmissions, respectively.

Similar performance patterns are found for both the aggregated migration time and VM downtime for the algorithms across various values of $SDRsc$ (Figure \ref{chap5-fig-perf-SDRsc-2}(b) and Figure \ref{chap5-fig-perf-SDRsc-2}(c)). Since both migration time and VM downtime are proportional to VM memory size, the above-mentioned explanation for migration data transmission metric also applies to these performance metrics. On average, compared to FFDL1 and MMDVMC, AMDVMC requires 77\% and 55\% less aggregated migration time and 79\% and 59\% less aggregated VM downtime, respectively, across all VM workload ranges. The aggregated network cost for both FFDL1 and MMDVMC algorithms increases sharply (Figure \ref{chap5-fig-perf-SDRsc-2}(d)) with the increase of $SDRsc$ since network cost is proportional to the migration data transmission and both FFDL1 and MMDVMC are network cost-unaware algorithms. AMDVMC, being network cost-aware, incurs 78\% and 68\% less cost than do FFDL1 and MMDVMC, respectively. 

\begin{figure}[!t]
\centering
\includegraphics[scale=0.45, trim=-1cm 11.5cm 0.2cm .6cm]{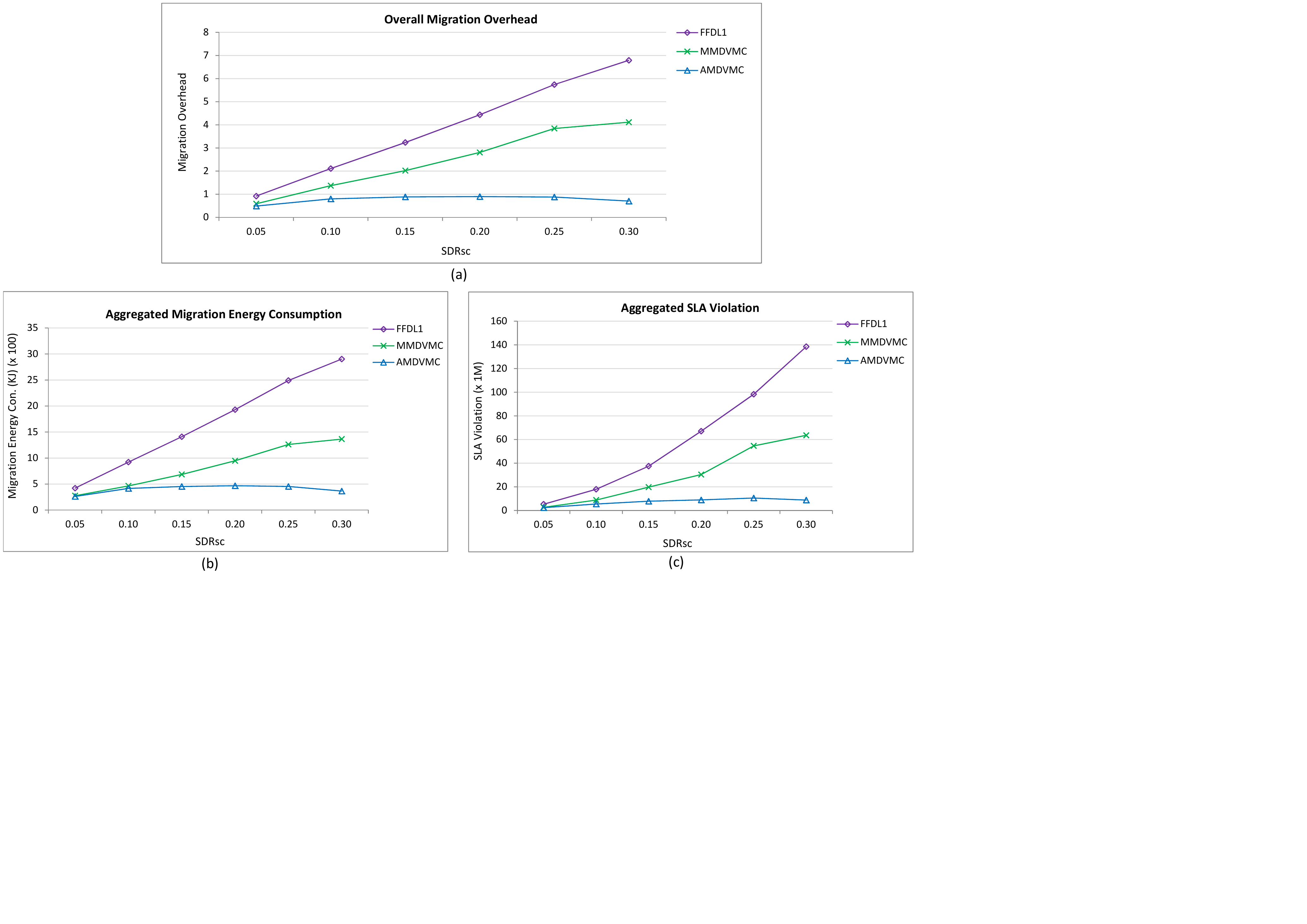}
\caption{Performance of the algorithms with increasing $S\!D\!Rsc$: (a) Overall Migration Overhead, (b) Aggregated Migration Energy Consumption, and (c) Aggregated SLA Violation.}
\label{chap5-fig-perf-SDRsc-3}
\end{figure}

The uniform migration overhead (Eq. \ref{chap5-eq-mo-2}) for all the algorithms is presented in Figure \ref{chap5-fig-perf-SDRsc-3}(a) and in summary, AMDVMC incurs 73\% and 57\% less migration overhead compared to FFDL1 and MMDVMC, respectively. Figure \ref{chap5-fig-perf-SDRsc-3}(b) and Figure \ref{chap5-fig-perf-SDRsc-3}(d) present the estimate of aggregated migration energy consumption and SLA violation due to consolidation decisions across various $SDRsc$ values. Since migration energy consumption and SLA violation are proportional to migration data transmission and VM migration time, respectively, these two performance metrics display similar performance patterns to those of Figure \ref{chap5-fig-perf-SDRsc-2}(a) and Figure \ref{chap5-fig-perf-SDRsc-2}(c), respectively. On average, compared to FFDL1 and MMDVMC, AMDVMC requires 68\% and 40\% less aggregated migration energy consumption, and 79\% and 58\% less aggregated SLA violation, respectively.

In view of the above results and analysis, it can be concluded that with the gradual increase of the diversification of workloads ($SDRsc$), the power consumption and resource wastage of the data center slowly increase for both FFDL1 and AMDVMC, whereas these metrics decrease for MMDVMC. However, all the cost factors increase rapidly for both FFDL1 and MMDVMC with the increase of workload diversification, while these factors remain largely steady for AMDVMC across workload variations. When compared to the migration-aware MMDVMC, the proposed AMDVMC algorithm outperforms MMDVMC significantly for both gain and cost factors. On the other hand, the migration-unaware FFDL1 algorithm achieves higher efficiency on the power consumption and resource wastage than AMDVMC, however this is achieved at the cost of very high migration overhead.

\subsection{AMDVMC Decision Time}
\label{chap5-sec-amdvmc-dec-time}
\begin{figure}[!t]
\centering
\includegraphics[scale=0.5, trim=0cm 17.3cm 2cm 2cm]{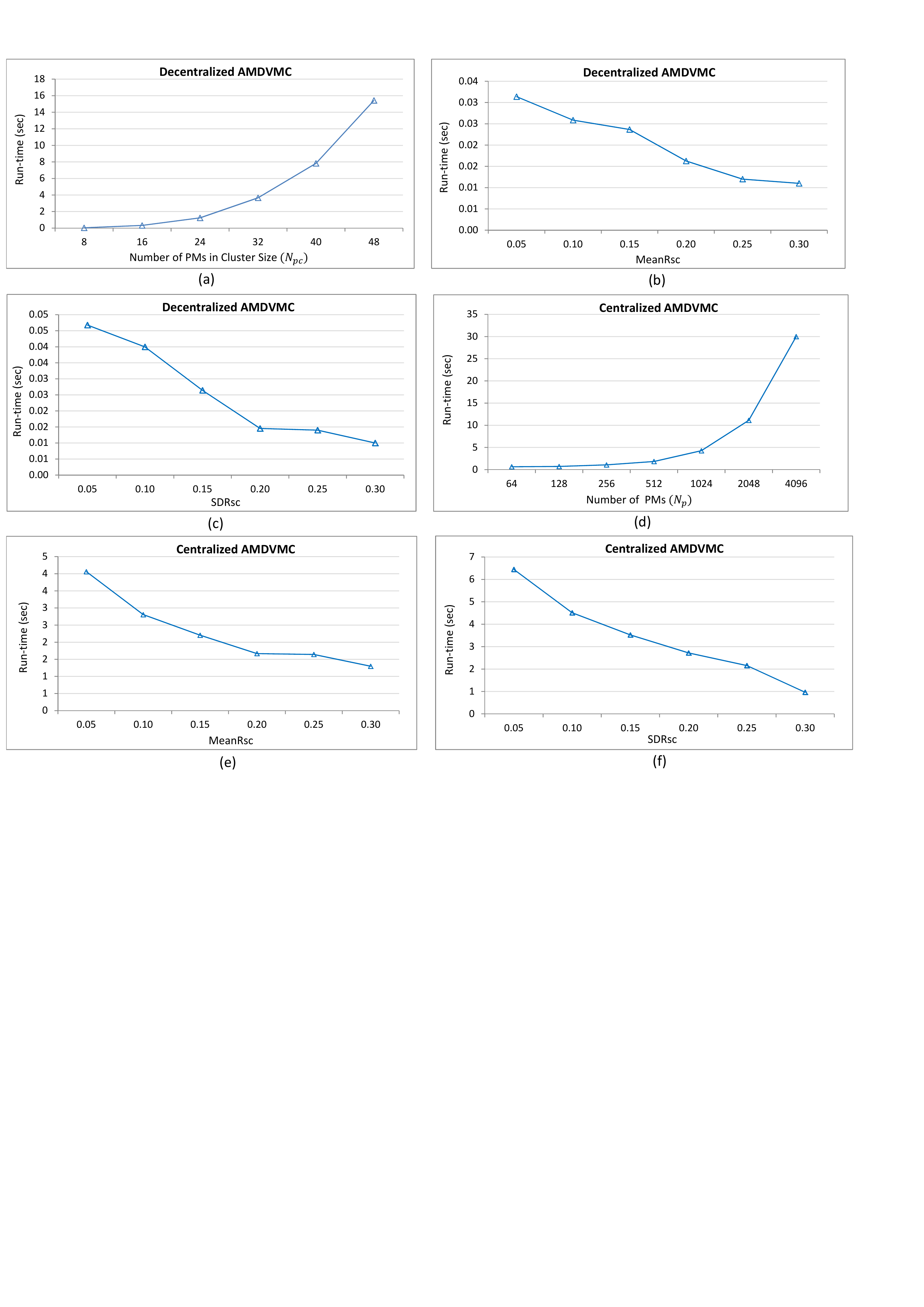}
\caption{AMDVMC's VM consolidation decision time for decentralized (a-c) and centralized (d-f) implementations while scaling PM cluster size ($N_{pc}$), Mean of VM resource demand ($M\!eanRsc$), and Diversification of VM workload ($S\!D\!Rsc$).}
\label{chap5-fig-perf-decision-time-1}
\end{figure}
This part of the experiment was conducted in order to assess the feasibility of the proposed AMDVMC algorithm for performing offline, dynamic VM consolidation for data center environments discussed in the problem statement (Section \ref{chap5-sec-dyn-vm-con}). As presented in Section \ref{chap5-sec-hierar-dec-dvmc-frame}, scalability of the proposed dynamic VM consolidation is ensured by running the VM consolidation operation under the proposed hierarchical, decentralized framework where each cluster controller is responsible for generating VM consolidation decisions for its respective PM cluster. Therefore, when implemented using the decentralized framework where the proposed AMDVMC dynamic VM consolidation algorithm is executed by the cluster controllers separately and simultaneously for their respective PM clusters, it is the cluster size ($N_{pc}$) that has a potential effect on the solution computation time rather than the total number of PMs in the data center. Figure \ref{chap5-fig-perf-decision-time-1}(a) shows AMDVMC's decision time for cluster sizes between 8 and 48. It can be observed that the decision time increases smoothly and non-linearly with the cluster size, each time doubling the time for an additional 8 PMs in the cluster, even though the search space grows exponentially with $N_{pc}$. For a cluster of size 48, the decision time is around 15.4 seconds which is quite a reasonable run-time for an offline algorithm. 

Figure \ref{chap5-fig-perf-decision-time-1}(b) and  Figure \ref{chap5-fig-perf-decision-time-1}(c) show the solution computation time while scaling the mean ($MeanRsc$) and standard deviation ($SDRsc$) of VM resource demand (in a similar way done in Section \ref{chap5-sec-scale-dc-size} and Section \ref{chap5-sec-scale-mean-rsc-dem}) for cluster size $N_{pc} = 8$. It can be observed from the figures that, in both instances, the decision time reduces with the increase of $MeanRsc$ and $SDRsc$. This is due to the fact that, in these instances, the number of VMs in the data center ($N_{v}$) declines with the increase of $MeanRsc$ and $SDRsc$ (Table \ref{chap5-tab-eval-Nv-MeanRsc-1} and Table \ref{chap5-tab-eval-Nv-SDRsc-1}), which reduces the solution computation time. In summary of these two cases, AMDVMC requires at most 0.05 second for computing consolidation plans. Therefore, when implemented using the proposed hierarchical, decentralized framework, it can be concluded that the proposed AMDVMC algorithm is a fast and feasible technique for offline, dynamic VM consolidation in the context of large-scale data centers.

In order to assess the time complexity of AMDVMC for scenarios where decentralized computation is not available, the solution computation time for a centralized system was also measured and analyzed. For this purpose, VM consolidation decisions for each of the PM clusters were computed in a centralized and single-threaded execution environment and the solution computation time for individual clusters were aggregated and reported in this evaluation. Figures \ref{chap5-fig-perf-decision-time-1}(d)-(f) show the average time needed by such a centralized implementation of the AMDVMC algorithm for producing dynamic VM consolidation plans for the various scaling factors. 

It can be observed from Figure \ref{chap5-fig-perf-decision-time-1}(d) that the AMDVMC solution computation time increases smoothly and non-linearly with the number of PMs in the data center ($N_{p}$). It is evident from the figure that for a medium sized data center comprising 1024 PMs, AMDVMC requires around 4.3 seconds for computing the VM consolidation plan whereas for the largest data center simulated in this experiment with 4096 PMs (i.e., several thousand physical servers), AMDVMC needs around 30 seconds.  Moreover, since AMDVMC utilizes the ACO metaheuristic which is effectively a multi-agent-based computation method, there is potential for parallel implementation \cite{Pedemonte2011a} of AMDVMC algorithm where individual ant agents can be executed in parallel in multiple Cloud nodes that can reduce the VM consolidation decision time significantly.

Furthermore, Figure \ref{chap5-fig-perf-decision-time-1}(e) and Figure \ref{chap5-fig-perf-decision-time-1}(f) show that the solution computation time of AMDVMC reduces with increasing $MeanRsc$ and $SDRsc$, respectively. This is also due to the above-mentioned fact that the number of VMs is reduced with increasing mean and standard deviation of VM resource demands accordingly to formulations \ref{chap5-eq-eval-Nv-MeanRsc-1} and \ref{chap5-eq-eval-Nv-SDRsc-1}, respectively. In summary of these two cases, AMDVMC requires at most 6.4 seconds for computing consolidation plans. Therefore, it can be concluded that, for centralized execution, the proposed AMDVMC algorithm is perfectly applicable for computing offline, dynamic VM consolidation plans for large-scale data centers.

\section{Conclusions and Future Directions}
\label{chap5-sec-conclusions}
Resource optimization has always been a challenging task for large-scale data center management. With the advent of Cloud Computing, and its rapid and wide adoption, this challenge has taken on a new dimension. In order to meet the increasing demand of computing resources, Cloud providers are deploying large data centers, consisting of thousands of servers. In these data centers, the run-time underutilization of computing resources is emerging as one of the key challenges for successful establishment of Cloud infrastructure services. Moreover, this underutilization of physical servers is one of the main reasons for power inefficiencies in data centers. Wide adoption of server virtualization technologies has opened opportunities for data center resource optimization. Dynamic VM consolidation is one of such techniques that helps in rearranging the active VMs among the physical servers in data centers by utilizing the VM live migration mechanism in order to consolidate VMs into a minimal number of active servers so that idle servers can be turned to lower power states (e.g., standby mode) to save energy. Moreover, this approach helps in reducing the overall resource wastage of running servers.

This paper has addressed a multi-objective dynamic VM consolidation problem in the context of large-scale data centers. The problem was formally defined as a discrete combinatorial optimization problem with necessary mathematical models with the goals of minimizing server resource wastage, power consumption, and overall VM migration overhead. Since VM migrations have non-negligible impacts on hosted applications and data center components, an appropriate VM migration overhead estimation mechanism is also suggested that incorporates realistic migration parameters and overhead factors. Moreover, in order to address the scalability issues of dynamic VM consolidation operations for medium to large-scale data centers, a hierarchical, decentralized consolidation framework was proposed to localize VM consolidation operations and reduce their impact on the data center network. Furthermore, based on ACO metaheuristic \cite{Dorigo1997}, a migration overhead-aware, multi-objective, dynamic VM consolidation algorithm (AMDVMC) was presented as a concrete solution for the defined run-time VM consolidation problem, integrating it with the migration overhead estimation technique and the decentralized consolidation framework. 

In addition, comprehensive simulation-based performance evaluation and analysis have also been presented that demonstrate the superior performance of the proposed AMDVMC algorithm over the compared migration-aware consolidation approaches across multiple scaling factors and several performance metrics, where the results show that AMDVMC reduces the overall server power consumption by up to 47\%, resource wastage by up to 64\%, and migration overhead by up to 83\%. Last but not least, the feasibility of applying the proposed AMDVMC algorithm as an offline dynamic VM consolidation technique in terms of decision time has been demonstrated by the performance evaluation, where it is shown that the algorithm requires less than 10 seconds for large server clusters when integrated with the proposed decentralized framework and a maximum of 30 seconds for large data centers when executed in centralized mode.

\subsection{Future Research Directions}
Cloud Computing, being a very dynamic environment, is rapidly evolving and opening new directions for further research and improvement. Moreover, VM management is a broad area of research, in particular in the context Cloud Computing. Based on the insights gained from the research presented in this paper, the following open research challenges are identified:

\begin{itemize}
\item Cloud environments allow their consumers to deploy any kind of applications in an on-demand fashion, ranging from compute intensive applications, such as High Performance Computing (HPC) and scientific applications, to network and disk I/O intensive applications, such as video streaming and file sharing applications. Co-locating similar kinds of applications in the same physical server can lead to resource contentions for some types of resources, while leaving other types of resources underutilized. Moreover, such resource contention will have adverse effect on application performance, thus leading to Service Level Agreement (SLA) violations and profit minimization. Therefore, it is important to understand the behavior and resource usage patterns of the hosted applications in order to efficiently place VMs and allocate resources to the applications. Utilization of historical workload data and application of appropriate load prediction mechanisms need to be integrated with dynamic VM consolidation techniques in order to minimize resource contentions among applications, and increase resource utilization and energy efficiency of the data centers.

\item Incentive-based VM migration and consolidation is yet another direction for future research in the area of Cloud Computing. Appropriate incentive policy can be formulated for Cloud consumers to trade off between the SLA violation and the provided incentive, which in turn will motivate Cloud providers to optimize infrastructure resources by specialized VM placement, migration, and consolidation strategies with the goal of saving energy and improving resource usage. 

\item Widespread use of virtualization technologies, high speed communication, increased size of data and data centers, and above all, the broad spectrum of modern applications are opening new research challenges in network resource optimization. Appropriate combination and coordination of the online and offline VM placement and migration techniques with the goal of efficient network bandwidth management is one of the key areas for future research.
\end{itemize}

\newpage

\bibliography{D:/Sunny/Dropbox/monash/bibtex-references/references-all}
\bibliographystyle{wileyj}

\end{document}